%% file: TIC.tex
\documentclass[10pt,twocolumn,notitlepage,aps,pra,showpacs,superscriptaddress,
floatfix,nofootinbib,longbibliography]{revtex4-1}

\AtBeginDocument{%
    \newwrite\bibnotes
    \def\bibnotesext{Notes.bib}
    \immediate\openout\bibnotes=\jobname\bibnotesext
    \immediate\write\bibnotes{@CONTROL{REVTEX41Control}}
    \immediate\write\bibnotes{@CONTROL{%
    apsrev41Control,author="08",editor="1",pages="1",title="0",year="1"}}
     \if@filesw
     \immediate\write\@auxout{\string\citation{apsrev41Control}}%
    \fi
}%

\usepackage[colorlinks=true,citecolor=blue,urlcolor=magenta]{hyperref}
\usepackage[utf8]{inputenc}
\usepackage[T1]{fontenc}
\usepackage{amsmath, amsthm, amssymb,amsfonts}
\usepackage{graphicx}
\usepackage{dcolumn}
\usepackage{bm}
\usepackage{bbm}
\usepackage{mathtools}
\usepackage{comment}
\usepackage{color}
\usepackage{diagbox}
\usepackage{tabularx}

\usepackage{graphicx}
\usepackage{amssymb}
\usepackage{amsmath}
\usepackage{amsthm}
\usepackage{color}
\usepackage{psfrag}
\usepackage{epsfig}
\usepackage{bbm}
\usepackage{bm}
\hypersetup{breaklinks}

\input{MischaCommandsV2}

\newcommand{\floor}[1]{\left\lfloor #1 \right\rfloor}
\newcommand{\ceil}[1]{ \left\lceil #1  \right\rceil }

\newcommand{\xx}{\mathbf x}
\newcommand{\yy}{\mathbf y}
\newcommand{\zz}{\mathbf z}
\newcommand{\LL}{\mathcal L}

\newcommand{\II}{\mathcal I}
\newcommand{\ZZ}{\mathcal Z}

\newcommand{\td}{\widetilde }

\newcommand{\mean}[1]{\langle #1 \rangle}
\newcommand{\id}{{\mathbbm{1}}} %Alternatively, can use \mathds{1} with \usepackage{dsfont}. Alternativerly use \mathbbm{1} with \usepackage{bbm}.

\providecommand{\openone}{{\mathbbm 1}}

\newcommand{\tr}{tr}

\newcommand{\ket}[1]{| #1 \rangle}

\newcommand{\de}{\partial}
\newcommand{\bra}[1]{\langle #1 |}

\newcommand{\ketbra}[1]{| #1 \rangle\langle #1 |}

\begin{document}
\title{Ticking-clock performance enhanced by nonclassical temporal correlations}
\author{Costantino Budroni}
\thanks{All authors contributed equally.}
\affiliation{Faculty of Physics, University of Vienna, Boltzmanngasse 5, 1090 Vienna, Austria}
\email{costantino.budroni@univie.ac.at}
\affiliation{Institute for Quantum Optics and Quantum Information (IQOQI), Austrian Academy of Sciences, Boltzmanngasse 3, 1090 Vienna, Austria}\email{giuseppe.vitagliano@oeaw.ac.at}

\author{Giuseppe Vitagliano}
\thanks{All authors contributed equally.}
\affiliation{Institute for Quantum Optics and Quantum Information (IQOQI), Austrian Academy of Sciences, Boltzmanngasse 3, 1090 Vienna, Austria}

\author{Mischa P. Woods\,}
\thanks{All authors contributed equally.}
\affiliation{Institute for Theoretical Physics, ETH Zurich, Switzerland}
\email{mischa@phys.ethz.ch}

%\date{\today}

\begin{abstract}  
We investigate the role of nonclassical temporal correlations in enhancing the performance of ticking clocks in a discrete-time scenario. We show that the problem of optimal models for ticking clocks is related to the violation of Leggett-Garg-type temporal inequalities formulated in terms of, possibly invasive, sequential measurements, but on a system with a bounded memory capacity. 
Ticking clocks inspire the derivation of a family of temporal inequalities showing a gap between classical and quantum correlations, despite involving no input.
We show that quantum ticking-clock models achieving accuracy beyond the classical bound are also those violating Leggett-Garg-type temporal inequalities for finite sequences and we investigate their continuous-time limit. Interestingly, we show that optimal classical clock models in the discrete-time scenario do not have a well-defined continuous-time limit, a feature that is absent in quantum models.
%\\
%\noindent PACS numbers:
\end{abstract}

\maketitle
\section{Introduction} 
The long-standing problem of understanding time in quantum theory has recently acquired renewed interest from the perspective of an operational definition of time and the practical and fundamental limitations to its measurement or even its definition \cite{ErkerPRX2017,Woods2018,RaLiRe15,ATGRandomWalk,Woods2021,SchwarzhansPRX2021,WendelPRL2020}.
At the same time, quantum information processing tasks involving temporal correlations and sequential operations are ubiquitous: from random access codes \cite{Wiesner1983,Ambainis1999, Ambainis2002,BowlesPRA2015, 
		AguilarPRL2018, Miklin2020}, to dimension witnesses \cite{GallegoPRL2010, BrunnerPRL2013, GuehnePRA2014, BudroniPRL2014, SchildPRA2015, Spee2020, Sohbi2021}, communication \cite{BrierleyPRL2015, Zukowski2014,Pironio_comm2003, Montina_comm2016}, simulations of  contextuality~\cite{KleinmannNJP2011,Fagundes2017} (see also \cite{Context_review}), purity certification \cite{Spee2019},  simulation of stochastic processes~\cite{GarnerNJP2017,Elliott2018,Elliott2019}, and memory asymmetry~\cite{ThompsonPRX2018}. 
		
		 The first approach to the study of temporal correlations dates back to Leggett and Garg~\cite{LeggettPRL1985}, who defined classical correlations from two assumptions: {\it macrorealism per-se} (MR), i.e., the existence of a definite value for a physical quantity at any time, and {\it noninvasive measurability} (NIM), i.e., the possibility of measuring such a quantity without altering its  value. From this, they derived the so-called {\it Leggett-Garg inequalities} (LGI), which were tested in a wide variety of physical systems (see the review \cite{EmaryRPP2014} and Refs.~\cite{RobensPRX2015, ZhouPRL2015, WangPRA2016, KneeNC2016, Formaggio2016, HuffmanPRA2017, KatiyarNJP2017, WangPRA2018, MajidyPRA2019} for more recent experiments). A major challenge is the {\it clumsiness loophole}  \cite{WildeMizel2012,HalliwellPRA2016,EmaryPRA2017,KneeNC2016,MajidyPRA2019,UolaPRA2019}: the possibility that the NIM condition is violated due to an imperfect, or clumsy, measurement rather than  quantum effects. 
This strong restriction on allowed operations makes it difficult to discuss in terms of LGI quantum advantages in information-theoretic tasks involving sequential operations: Any classical device with an internal memory that is updated sequentially would violate NIM. To overcome this problem, a new framework has been proposed~\cite{BudroniNJP2019}, which relaxes NIM assumption to that of bounded internal memory: The operations are allowed to be invasive, but they can modify an internal memory of at most $n$ bits; NIM is recovered for the special case $n=0$.
  
Here, we address the problem of distinguishing quantum from classical clocks based on temporal correlations in the spirit of Leggett-Garg. We start from a discrete-time structure and then analyze its continuous-time limit. We first show how the notion of accuracy of a ticking clock gives rise to a family of temporal inequalities and prove analytically the bound for the bit case. Such inequalities, involving first and second moments of the ticks distribution, however, need infinitely long sequences. We, then, derive another family of inequalities, which discriminates classical and quantum systems for finite sequences. For the finite sequences, experimental tests can be easily designed, in analogy with standard LGI tests. Finally, we show that quantum models that achieve an accuracy above the classical bound are also those that violate such inequalities. Interestingly, optimal classical clocks do not always have a well-defined continuous limit, in contrast with quantum clocks, which are readily extended to the continuous-time limit in all cases.

%%%%%%%%%%%%%%%%%%%%%%%%%%%%%%%%%%%%%%%%%%%%

%%%%%%%%%%%%%%%%%%%%%%%%%%%%%%%%%%%%%%%%%%%%

\section{Preliminary notions} 
To relax the constraint of NIM, we adopt the framework developed in \cite{Hoffmann2018,BudroniNJP2019,SpeeNJP2020}, based on the notion of a {\it finite-state machine}: a box that 
receives an input $x\in \mathcal{X}$, produces an output $a\in \mathcal{A}$, then  receives another input $y$ and produces another output $b$, and so on.  The classical version of a finite-state machine of dimension $d$ can be compactly represented as
\begin{equation}\label{eq:defPclass}
p(ab|xy)= \pi T(a|x) T(b|y) \eta .
\end{equation}
Here $\pi\in \mathbb{R}^d$  is a (row) vector with components $[\pi]_j \geq 0$ representing the probability of being in the state $j$, and satisfying $\sum_j [\pi]_j=1$, the matrix $T(o|i)$, associated with output $o$ and input $i$, is $d\times d$ and substochastic, i.e., $[T(o|i)]_{jk}\geq 0$ and $\sum_k [T(o|i)]_{jk} \leq 1$, such that $\sum_o T(o|i)$ is a stochastic matrix, i.e.,  $\sum_{k,o} [T(o|i)]_{jk} = 1$, where each entry $[T(o|i)]_{jk}$ represents the probability of transitioning from the state $i$ to the state $j$ when the input $i$ is given and the output $o$ is observed. Finally, $\eta\in \mathbb{R}^d$ is the column vector $(1,\ldots,1)$, which provides a sum over all possible final states of the machine. Notice that the set of operations $\{T(o|i)\}_{o,i}$ is assumed to be fixed along the sequence, whenever the same pair of input-output is observed, the same state transformation is applied.

In contrast, in quantum mechanics, probabilities are given by
\begin{equation}\label{eq:quantumP}
p(ab|xy) := \tr[\II_{b|y} \circ \II_{a|x}(\rho)],
\end{equation}
where $\rho$ is the $d$-dimension initial state, $\{\II_{o|i}\}_o$ is the quantum instrument associated with the input $i$, i.e., $\II_{o|i}$ is a completely positive (CP) map for all $o$ and $\sum_{o} \II_{o|i}$ is a completely-positive trace-preserving (CPTP) map, and $\circ$ denotes composition. Similarly, $\{\II_{o|i}\}_{o,i}$ are assumed to be fixed along the sequence.

The classical case, extensively discussed in the literature (see, e.g., \cite{PazBook2003}), corresponds to the quantum case of states and operations diagonal in the same basis, as can be easily verified, e.g., via the Choi-Jamio\l{}kowski isomorphism. Classical and quantum models provide different correlations that can be distinguished via Leggett-Garg-type inequalities~\cite{BudroniNJP2019}. The finite memory constraint can be interpreted as a relaxation of Leggett and Garg's NIM assumption: Measurements are now allowed to be invasive, but up to a limited amount fixed by the system memory.
The constraint of bounded memory is fundamental in the investigation of nonclassical temporal correlations, since classical system with unbounded memory are able to simulate all temporal correlations~\cite{FritzNJP2010}, see also the discussion in \cite{Hoffmann2018,BudroniNJP2019}. 

In its simplest instance, a {\it ticking clock} is modeled by a physical system generating an output $1 =$``tick'' at some instants of time. In our model, time is discrete. Our clock models are given by finite-state machines without input (or, equivalently, with $\mathcal{X}=\{0\}$ consisting of only one symbol), where the dynamics is described by either transition matrices $\{T_0, T_1\}$  or a quantum instrument $\{ \II_0, \II_1\}$ associated with outputs $\mathcal{A}=\{0,1\}=\{{\rm ``tick'', ``no\ tick''}\}$.  We call this model a {\it discrete-time ticking clock}. The choice of no input is required for the clock to be {\it autonomous}, namely, not relying on some external input to generate the time signal.
In contrast to previous models, e.g., \cite{RaLiRe15}, we do not require a  so-called \emph{gear system} assigning the output-memory allocation, and our model closely resembles a discrete-time version of \cite{Woods2021}. This is the simplest measurement model one can imagine in the temporal scenario. Nevertheless, due to its nontrivial causal structure, this model already features observable differences between classical and quantum correlations, in contrast to the standard Bell and Leggett-Garg scenarios, in which an external input is needed.

%%%%%%%%%%%%%%%%%%%%%%%%%%%%%%%%%%%%%%%%%%%%

%%%%%%%%%%%%%%%%%%%%%%%%%%%%%%%%%%%%%%%%%%%%

\section{Accuracy of discrete-time clocks} 
A figure of merit of the quality of a ticking clock (for a clock that resets itself after each tick) is the {\it accuracy}~\cite{ErkerPRX2017}
\begin{equation}
R:=\frac{\mu^2}{\sigma^2},
\end{equation}
where $\mu$ is the mean time interval between ticks and $\sigma^2$ its variance. This quantity corresponds to the average number of  ticks that the clock can make before its uncertainty is larger than $\mu$. With this figure of merit, it has been shown that continuous-time quantum clocks can outperform classical clocks of the same dimension~\cite{Woods2018}.
For a discrete-time clock, defining by $p(L):= p(00\ldots 01)$ the probability of the first occurrence of the outcome ``1'' at time-step $L$, one defines the mean and the variance of such a distribution as $\mu := \sum_{L=1}^\infty L p(L)$ and $\sigma^2 := \sum_{L=1}^\infty (L-\mu)^2 p(L)$.
Using the arrow-of-time conditions~\cite{ClementePRL2016}, namely, the condition of no signaling from the future to the past, one can write 
\begin{equation}\label{eq:p000m}
\begin{split}
p(L)=
p(\!\!\!\underbrace{0\ldots0}_{L-1 \text{ zeros}}\!\!\!)-p(\underbrace{0\ldots0}_{L  \text{ zeros}}) := f(L-1)- f(L),
\end{split}
\end{equation}
where we defined the corresponding $f:\mathbb{N}\rightarrow \mathbb{R}$.
This expression has a convenient form that allows the employment of the notion of {\it Z-transform} to simplify the calculations.
This is defined for  $f:\mathbb{N}\rightarrow \mathbb{R}$ as
\begin{equation}
\ZZ[f](z):= \sum_{n=0}^\infty f(n)z^{-n} =: \td f(z) ,
\end{equation} 
where $z$ is a complex variable.
The $Z$-transform satisfies a series of properties, such as linearity, bijectivity,  or properties associated with shifts and derivatives that will be useful in the following (see, e.g.,~\cite{HowardBook}). In particular, we  use that 
\begin{align}
\label{eq:z_shiftm}
\ZZ[f(n+1)]&=z(\ZZ[f(n)]-f(0)), \\
\label{eq:z_shift2m}\ZZ[f(n)-f(n-1)] &= (1-1/z)\ZZ[f(n)],
\end{align}
which can be straightforwardly verified. Combining Eqs.~\eqref{eq:p000m} and \eqref{eq:z_shift2m}, we can write
\begin{equation}\label{eq:Q_f}
\begin{split}
Q(z):=\ZZ[p(L)] &= -(1-1/z)\ZZ[f(L)] \\
&= -(1-1/z)\tilde f(z).
\end{split}
\end{equation}
From the explicit expression $Q(z)=\sum_{L=0}^\infty p(L)z^{-L}$, one sees that $Q(z)$ is  also a generating function for the moments of the probability distribution $p(L)$, namely
\begin{align}
\label{eq:moments_qz2}\mu &= \sum_{L=1}^\infty L p(L) = -\left.\frac{\partial Q(z) }{\partial z} \right\rvert_{z=1}, \\ %= \tilde f(1),
 \sigma^2 &=  \sum_{L=1}^\infty L^2 p(L) - \mu^2 = \left. \frac{\partial^2 Q(z) }{\partial z^2} \right\rvert_{z=1} 
+ \left. \frac{\partial Q(z) }{\partial z} \right\rvert_{z=1} - \mu^2, 
\end{align}
which combined with Eq.~\eqref{eq:Q_f} give
\begin{equation}\label{eq:momentsfromf}
\mu = \tilde f(1), \qquad \sigma^2 = -\mu(\mu-1) - 2\tilde f^{\prime}(1),
\end{equation}
where $\tilde{f}':= \frac{\partial \tilde{f} }{\partial z}$.
Note that so far the calculations do not involve any particular model (i.e., classical or quantum) to compute $p(L)$.
In the following, we specialize first to the classical case, for which the linearity of the Z-transform can be fully exploited.

\subsection{Classical case}

For a classical automaton with transition matrices given by $(T_0,T_1)$ for the two outcomes ``0'' and ``1'', we have (see Eq.~\eqref{eq:defPclass})
\begin{equation}
 f_{\rm cl}(L)=\pi_0 T_0^L \eta ,
\end{equation}
where $\pi_0$ is the initial state.

 To compute $\td f_{\rm cl}(z)=\ZZ[f_{\rm cl}(L)]$ we need $\ZZ[T_0^L]$, which can be computed as follows.
Let us denote by $\pi(L)$ the state at the time step $L$ and by $T_0$ the transition matrix, we have the relation
\begin{equation}\label{eq:it_cond_ztr}
\begin{split}
\pi(0):=& \pi_0,\\
\pi(L+1)=& \pi(L) T_0 .
\end{split}
\end{equation}
By applying the $Z$-transform on both sides of the second equation and using Eq.~\eqref{eq:z_shiftm} and linearity, we obtain
\begin{equation}
\begin{aligned}
z \left( \td \pi(z)- \pi_0 \right) &= \td \pi(z) T_0,\\
\Rightarrow \td \pi(z) \left( \openone - z^{-1} T_0 \right) &= \pi_0,\\
\Rightarrow  \td \pi(z) = \ZZ[\pi(L)]= \pi_0 \ZZ[T_0^L] &= \pi_0 \left( \openone - z^{-1} T_0 \right)^{-1}.
\end{aligned}
\end{equation}
 This equation has solution iff $z$ is greater than the spectral range of $T_0$ (cf. Th.2.1 of Ref.~\cite{Seneta2006}). We then have
\begin{equation}\label{eq:qzclassicalm}
\begin{aligned}
Q(z)&= -(1-1/z)\ZZ[f_{\rm cl}(L)] = -(1-1/z)\pi_0\cdot \ZZ[T_0^L]\cdot \eta \\
&= -(1-1/z)\pi_0 \left( \openone - z^{-1} T_0 \right)^{-1} \eta.
\end{aligned}
\end{equation}

Thus, simplifying a bit the expression \eqref{eq:qzclassicalm} and transforming back, we obtain
\begin{equation}\label{eq:pLfromZm}
p(L)= \pi_0  \left(\frac 1 {2\pi i} \oint_C z^{L-1}(1-z)  \left( z\openone - T_0 \right)^{-1}  {\rm d} z \right) \eta ,
\end{equation}
where $C$ is a closed region inside the convergence region. Note that the matrix inside the square parentheses is nothing more than $T_0^{L-1}(\openone-T_0)$. Interestingly, the above integral representation resemble the spectral representation for Hermitian operators, where the matrix $\left( z\openone - T_0 \right)^{-1}$, called 
the {\it resolvent} of $T_0$,  plays a role similar to the spectral projections in computing a function of $T_0$, a polynomial in this case,  but for a matrix $T_0$ that is in general not Hermitian or even diagonalizable.

In turn, the integral in \eqref{eq:pLfromZm} can be calculated from the residues of the function 
\begin{equation}\label{eq:res_gen}
\begin{aligned}
z^{L-2}(1-z)\tilde f_{\rm cl}(z)&=z^{L-1}(1-z) \pi_0 \left( z\openone -  T_0 \right)^{-1} \eta. 
%\\
%&= \frac{z^{L-1}(1-z)}{\det(z\openone -  T_0)} \pi_0 \cdot {\rm adj}(z\openone - T_0) \cdot \eta ,
\end{aligned}
\end{equation}

The moments of $p(L)$ can, then, be calculated from Eq.~\eqref{eq:momentsfromf} as
\begin{equation}\label{eq:muqzcl}
\begin{aligned}
\mu &= \tilde f_{\rm cl}(1) =  \pi_0  \left( \openone -  T_0 \right)^{-1}\eta ,\\
\sigma^2 &= \mu(\mu-1)-2\tilde f_{\rm cl}^{\prime}(1) \\
&=\mu(1-\mu)+ 2\pi_0  \left( \openone -  T_0 \right)^{-1} T_0 \left( \openone - T_0 \right)^{-1}\eta \\
&=-\mu(\mu +1) +2\pi_0  \left( \openone -  T_0 \right)^{-2}\eta, %\label{eq:sigmaqzcl}
\end{aligned}
\end{equation}
since $\tilde f_{\rm cl}^\prime (z) = -\pi_0  \left( \openone - z^{-1} T_0 \right)^{-1} T_0 \left( \openone - z^{-1} T_0 \right)^{-1}\eta$ and where in the last line
we used the property that $[T_0,\left( \openone - z^{-1} T_0 \right)^{-1}]=0$.

So far, the derivations have been carried out for an arbitrary dimension. However, the actual calculation of the moments and the $p(L)$ 
become quickly very complicated as the dimension grows.
In the following, we provide the full solution for the most accurate classical clock model in the bit case. 

\subsubsection{Results in the bit case}

Let us consider here a general transition matrix for an automaton in $d=2$: 
\begin{equation}\label{eq:def_T0_const}
T_0=\left(
\begin{array}{cc}
a & b \\
c & d 
\end{array}
\right), 
\end{equation}
with  $a,b,c,d\geq 0 $ and $a+b, c+d \leq 1$. We can now directly compute the accuracy as a function of the model parameters $a,b,c,d$ via Eq.~\eqref{eq:muqzcl}. Note that since $p(L)$ is a linear function of $\pi$, by convexity we can choose the initial state to be pure, i.e., either $\pi_0=(1,0)$ or $\pi_0=(0,1)$. Up to a relabeling of the basis elements, we can fix the initial state to be $\pi_0=(1,0)$. 

Our goal here is to maximize the accuracy. However, it is easy to see that for $\mu = 2$ the accuracy can be infinity, as one can easily construct a bit clock that ticks every two time-steps.  More in general, one can see that accuracy of a clock can be infinite if the mean is smaller or equal than the dimension. A non-trivial solution can be obtained only if we fix $\mu > 2$, or $\mu> d$ in the general case. 

Interestingly, it is possible to solve analytically for a generic $\mu > 2$, using the Karush-Kuhn-Tucker (KKT) conditions \cite{CVXBook}, a generalization of Lagrange multiplier for inequality constraints. The solution is given by $b=2/\mu$, $a=d=1-b$, $c=0$. See Appendix~\ref{app:2_dim_cl} for the details.
As a solution we find that, as anticipated, for $\mu \leq 2$ one can obtain a perfect clock: $\sigma^2=0$ and $R=\infty$. For $\mu> 2$ the maximum accuracy is
\begin{equation}
	R_{\rm opt}=2\mu/(\mu-2) ,
\end{equation}
achieved by
\begin{equation}\label{eq:Ropt}
T_0=\left(
\begin{array}{cc}
q & 1-q \\
0 & q 
\end{array}
\right),\quad \mbox{with} \quad  q = 1-\frac{2}{\mu},
\end{equation}
which we call {\it one-way} or {\it discrete ladder clock} model. From this solution we can extract an inequality for two-state machines: 
\begin{equation}\label{eq:ti_mu}
R\leq 2\mu/(\mu-2) \ \stackrel{\mu > 2}{\Rightarrow} \ \mu ({\mu-2}) - 2 \sigma^2 \leq 0 .
\end{equation}
We have proven analytically the condition in Eq.~\eqref{eq:ti_mu} in the case of machine of dimension two, however, on the basis of an extensive numerical search, we expect to generalize to arbitrary dimension as
\begin{equation}\label{eq:conj1}
 \mu ({\mu-d}) - d \sigma^2 \leq 0.
\end{equation}
See the discussion in Sec.~\ref{sec:high_dim} and in Appendix~\ref{app:num_search}.

\subsection{Qubit clock beating the classical accuracy}

\begin{figure}[t]
\includegraphics[width=0.49\textwidth]{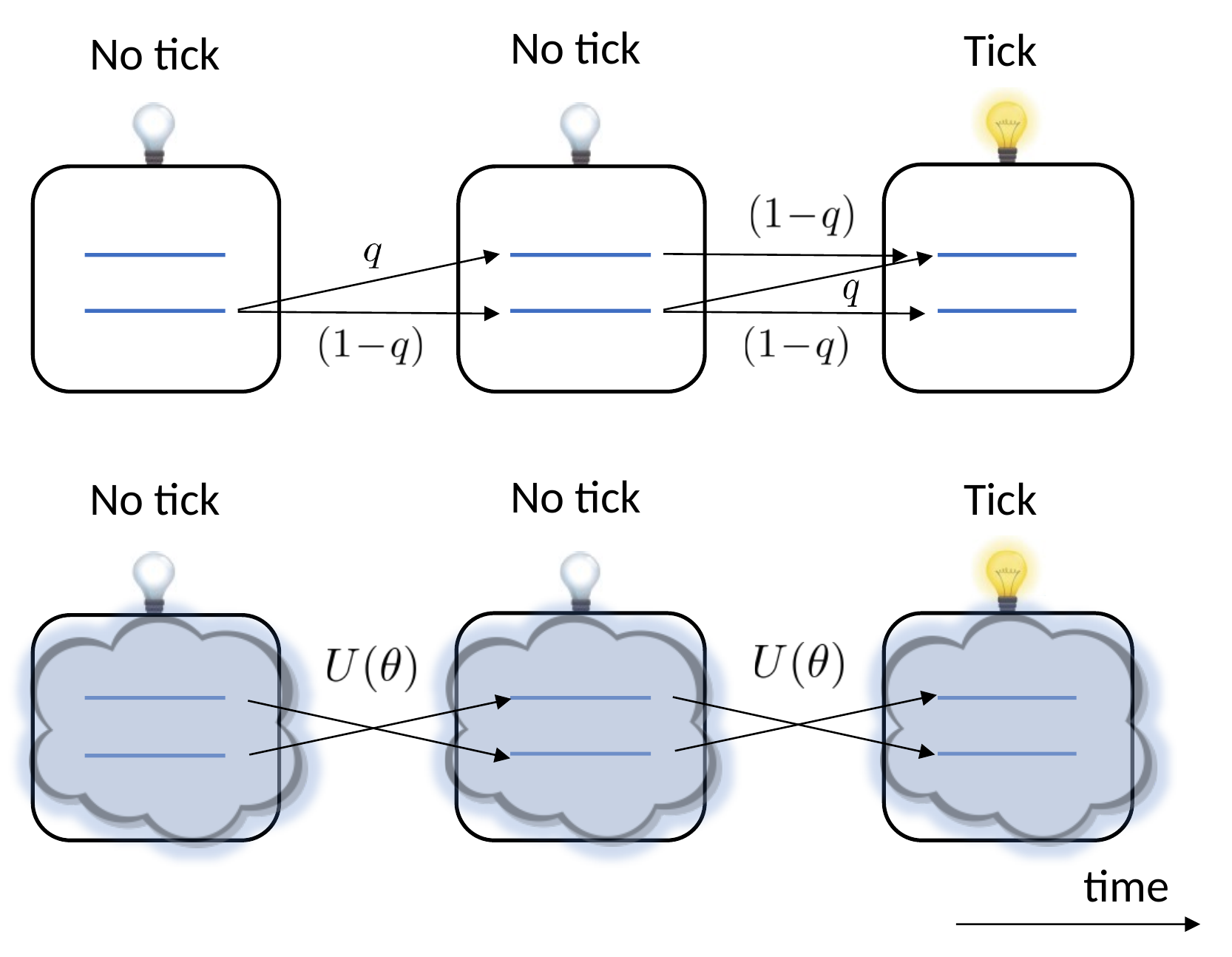}
\caption{\label{fig:q_vs_c_clock} The classical discrete ladder clock model versus its quantum counterpart. In the classical model,  the system transition with probability $q$ to the next state or remains with probability $(1-q)$. The probability of ``tick'' is zero for all states, except the last one, where it is $(1-q)$. In the quantum model, the ``no-tick'' instrument is of the form $\mathcal I_0= \exp(iH\theta)\sqrt{E_0}$ and the initial state is a superposition of the eigenbasis of $H$. Similarly to the classical case, the probability of ``tick'' is non-zero only in the last state (in $E_0$ eigenbasis), however the system is coherently rotated, by an angle $\theta$ at each step, such that all this intermediate states have a non-zero overlap with the last state. }
\end{figure}

Inspired by the classical cyclic model, we construct a qubit model able to violate Eq.~\eqref{eq:ti_mu}, for any $\mu > 2$, in terms of only two real parameters $q$ and $u$.
The model is described by a single Kraus operator for the ``no-tick'' output, namely
$K_0=U_0 \sqrt{E_0}$, with 
\begin{equation}\label{eq:q_modelEU}
E_0=\left(
\begin{array}{cc}
1 & 0 \\
0 & q^2
\end{array}
\right), \qquad U_0=\left(
\begin{array}{cc}
\sqrt{u} & \sqrt{1-u} \\
-\sqrt{1-u} & \sqrt{u} 
\end{array}
\right),
\end{equation}
see also Fig.~\ref{fig:q_vs_c_clock}.
Analogously to the classical case, $\mu$ and $\sigma^2$ can be computed  via the $Z$-transform method, where now the moment generating function is
\begin{equation}
Q(z) = -(1-1/z) \mathcal Z[f_{\rm q}(L)] = -(1-1/z) \tilde f_{\rm q}(z) ,
\end{equation}
and from Eq.~\eqref{eq:quantumP} we find
\begin{equation}
f_{\rm q}(L):=\bra{\psi} (K_0^\dagger)^L K_0^L \ket{\psi} ,
\end{equation}
again considering simply a pure state by a similar convexity argument.
Note that in this case we cannot simplify much further the problem with the aid of the Z-transform, due to the fact that the expression is quadratic in the parameters of $\ket{\psi}$ rather than linear, thus rendering a full optimization more complicated in this case. 

Nevertheless, for the particular model defined by Eq.~\eqref{eq:q_modelEU}, we can optimize the parameters $q$ and
$u$ to find the optimal violation of  Eq.~\eqref{eq:ti_mu}. To do so, we minimize 
$\sigma^2$ over $q$ and $u$ for fixed $\mu(q,u)$. Full details are presented in Appendix~\ref{app:q_models}.

The result of this optimization provides the values
$u={2q}/( {1+q^2})$, and $q= \left( 1 - \frac 2 {\mu} \right)$. This solution gives
for the l.h.s. of Eq.~(\ref{eq:ti_mu}) (for $\mu > 2$)
\begin{equation}
\mu ({\mu-2}) - 2 \sigma^2 = \frac{\mu^2 (\mu-2)^2}{2(\mu-1)^2}  > 0 ,
\end{equation}
and $R\rightarrow 4\mu(\mu-1)^2/\left\lbrace(\mu-2)\left[\mu(\mu-2)+2\right]\right\rbrace$, giving  $R= 4= d^2$ for $\mu \rightarrow \infty$. 

This model closely resembles a discrete version of the quasi-ideal clock presented in~\cite{Woods2021,Woods2018,WSO16}, in that it has a single Kraus operator with free evolution (i.e., $U_0$) and measurement (i.e., $E_0$) having eigenbases related by the (discrete in our case) Fourier transform.

%%%%%%%%%%%%%%%%%%%%%%%%%%%%%%%%%%%%%%%%%%%%

%%%%%%%%%%%%%%%%%%%%%%%%%%%%%%%%%%%%%%%%%%%%

\begin{figure*}[ht]
\includegraphics[width=0.32 \linewidth]{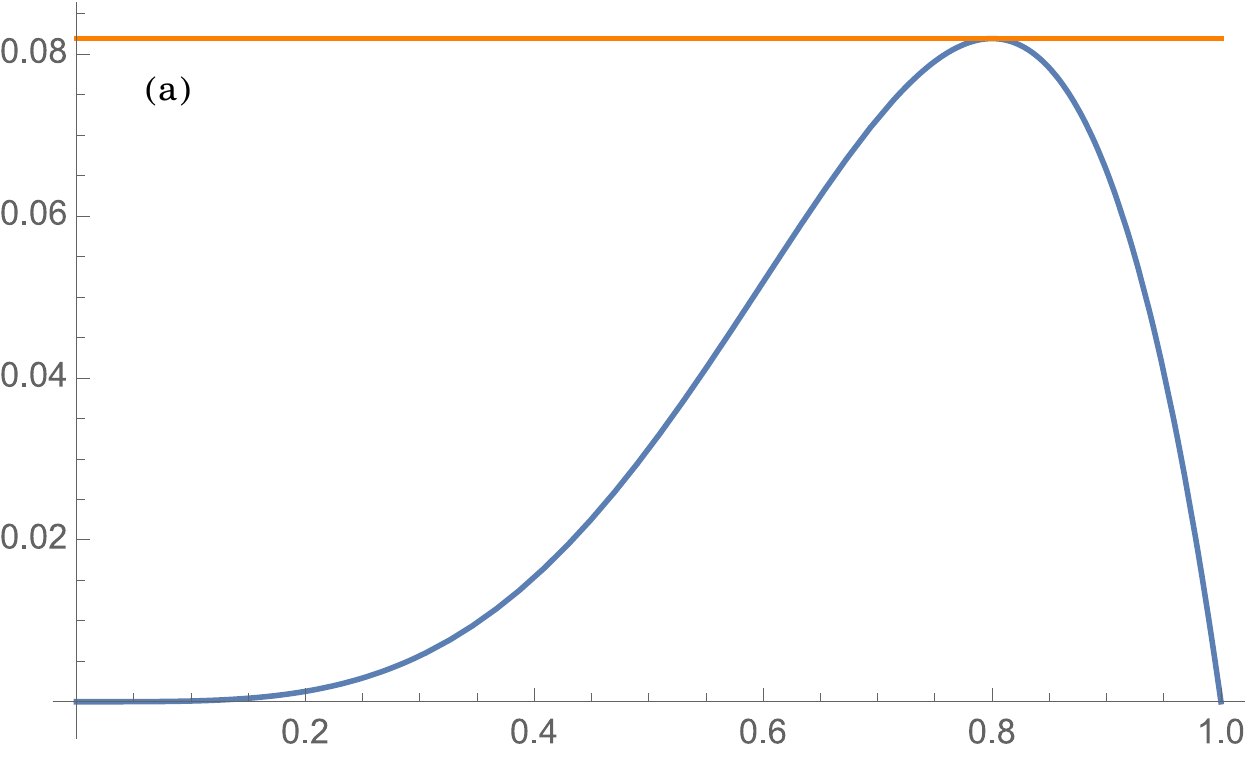} 
\includegraphics[width=0.32 \linewidth]{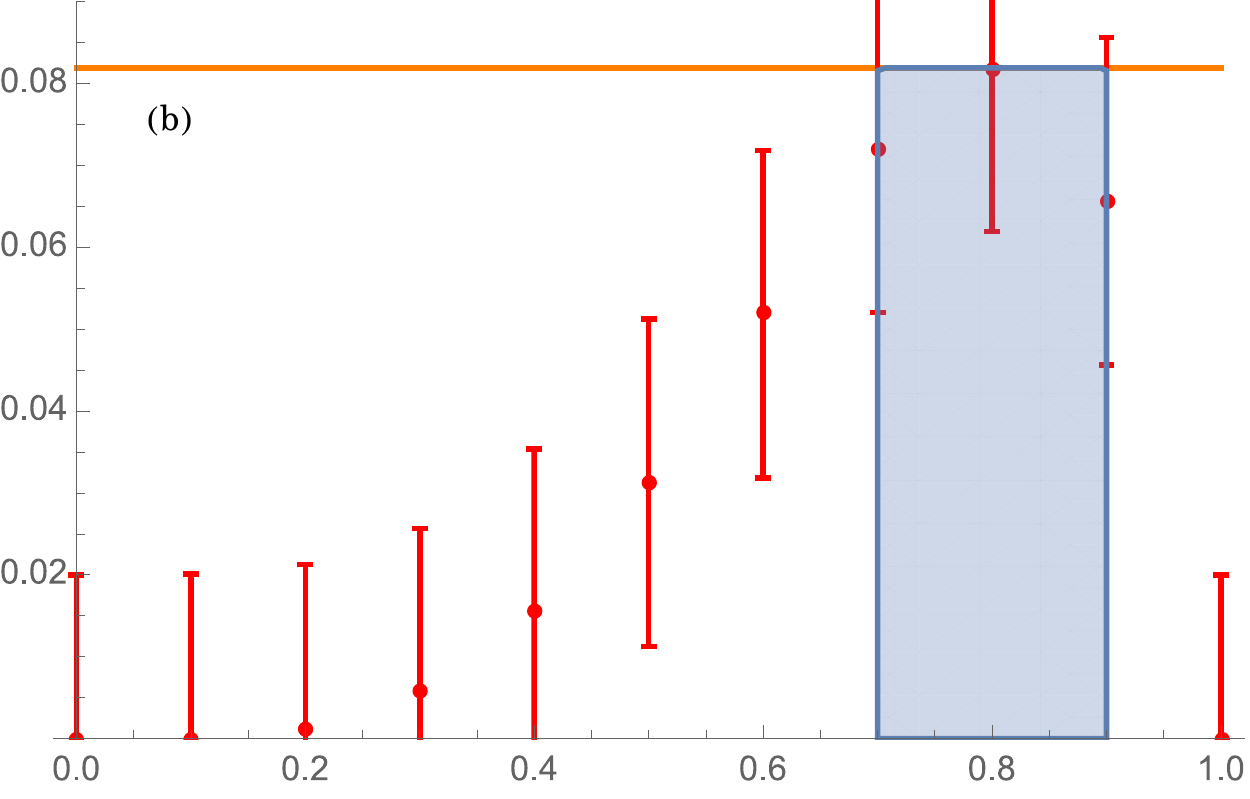}
\includegraphics[width=0.32 \linewidth]{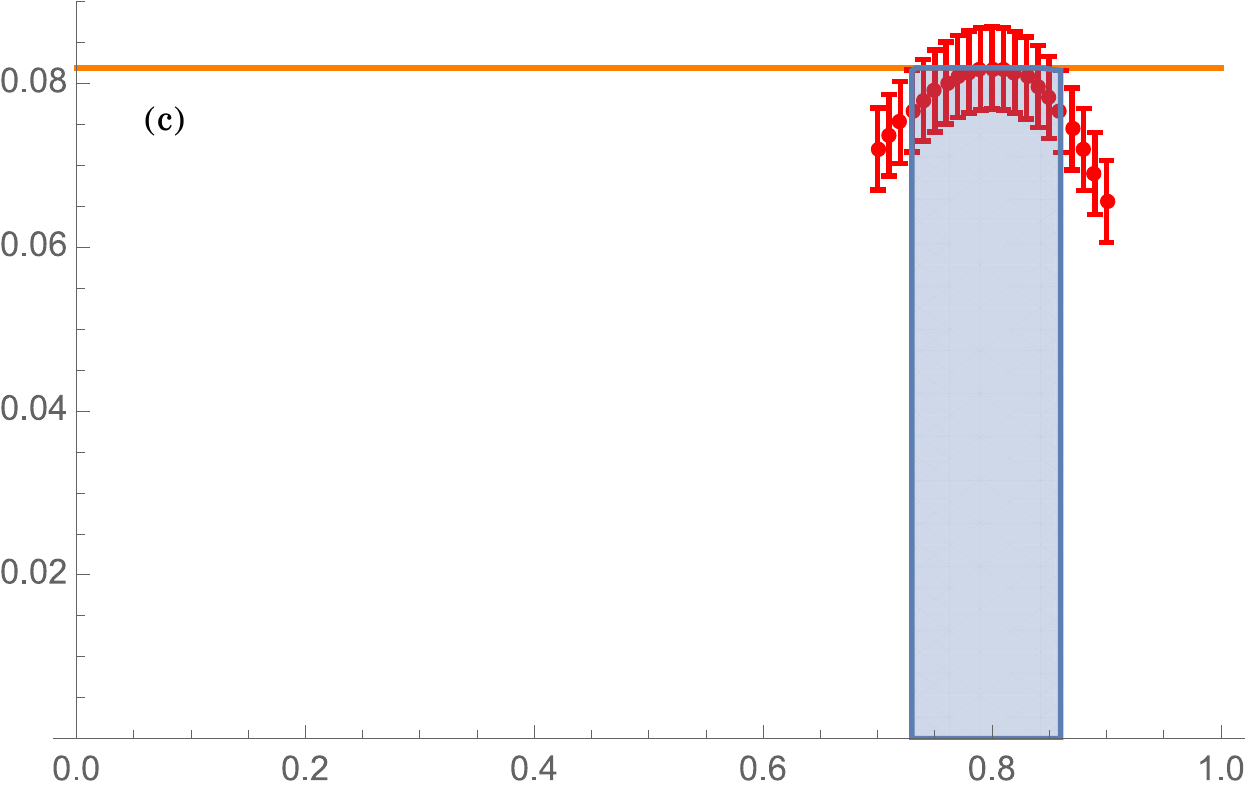}
\caption{\label{fig:iter} Pictorial description of the iterative grid-search method in one dimension. (a) Function we want to maximize (blue curve) and estimated maximum (orange line). (b) Computation of the function on a lattice with corresponding error (red dots and bars). According to the error estimate, only the points in the blue region are able to reach the estimated maximum. (c) As a subsequent step, the computation is performed on a refined lattice, with refined error, corresponding to the blue region in (b). Only a subset of those points are able to violate the maximum (new blue region), on which a new refined lattice is defined. These steps are repeated until the desired error is reached. Obviously, the same algorithm works even without an initial estimate of the maximum (orange line), by simply  using the maximum on the lattice as an estimate}
\end{figure*}

\section{Temporal inequalities for finite-length sequences}

\subsection{Bit case}

Equation \eqref{eq:ti_mu} can be thought of as Leggett-Garg-type inequality  providing a witness for nonclassical temporal correlations, w.r.t. the finite-state machine framework. In contrast to usual Leggett-Garg-type inequalities, however, it involves an infinite sequence of measurements. Inspired by the above construction, we  find a family of temporal inequalities for finite length $L$:
\begin{equation}\label{eq:def_temp_ine}
p(L) = p(00\ldots01)\leq \Omega_{d=2,L}^{\rm C} ,
\end{equation} 
where $\Omega_{d=2,L}^{\rm C}$ is an upper bound valid for all classical finite-state machines with $d=2$. By optimizing the above probability a second model other than the ladder clock arises, which we call the {\it cyclic} model:
\begin{equation}
T_0=\left(
\begin{array}{cc}
0 & 1 \\
r & 0 
\end{array}
\right) ,\qquad \text{with } r = 1- \ceil{\frac{L}{2}}^{-1},
\end{equation}
where $\ceil{\cdot}$ is the ceiling function. Still, the ladder model gives the optimum for certain specific L, and we similarly obtain $q = 1-\frac{2}{L}$ as the optimal corresponding value of the model parameter. More details can be found in Appendix~\ref{app:class_mod}, where we also study generalizations of such model for higher dimensions.

For the case $d=2$ we are indeed able to prove, up to $L=20$, an analytical upper bound to $\Omega^{\rm C}_{2,L}$, with a gap of $10^{-4}$ w.r.t. the maximum achieved by the cyclic and the one-way models. Such precise upper bound can be obtained via grid-search methods, exploiting the fact that the probability function depends only on four parameters, i.e., the entries of $T_0$ defined in Eq.~\eqref{eq:def_T0_const}. Let us give a brief sketch of the main ideas. Given a function $f:\mathbb{R}^D\rightarrow \mathbb{R}$ with proper continuity properties, we can write it in the lowest order in Taylor expansion with Lagrange remainder as
\begin{equation}\label{eq:taylor}
f(\xx) = f(\yy) + \nabla f(\zz) \cdot( \xx -\yy),
\end{equation}
 with $\zz = \lambda \xx + (1-\lambda) \yy$  for some $0\leq \lambda\leq 1$. 
In our case, $f$ represents the probability $p(L)$ as a function of $a,b,c,d$ in Eq.~\eqref{eq:def_T0_const}, on 
a compact subset $\mathcal{B}$ defined by positivity and normalization constraints. First, notice that $f$ and 
$\|\nabla f\|$ are continuous functions and $\mathcal{B}$ is compact, so they have a maximum in $\mathcal{B}$. 
Thus, if instead of evaluating $f$ on $\xx$, we evaluate on some other point $\yy$, e.g., belonging to a (hyper)cubic 
lattice of step $\delta$,  $\mathcal{L}_\delta := (\delta\mathbb{Z}^D)\cap \mathcal{B}$, we can bound the error 
in this approximation as the maximum of the gradient times the maximal distance between $\mathcal{B}$ and 
$\mathcal{L}_\delta$, namely, 
\begin{equation}\label{eq:lower_upper_f_m}
\max_{\xx \in \mathcal{B}} f(\xx)\leq \max_{\xx \in \LL_\delta} f(\xx) + \frac{\delta\sqrt{D}}{2} \max_{\xx \in \mathcal{B}} \|\nabla f(\xx)\|,
\end{equation}
where we used that $\delta\sqrt{D}/2$ is the maximal distance between $\mathcal{B}$ and $\mathcal{L}_\delta$. A peculiar property of our probability function $f$ is that also the components of its gradient can be interpreted as probabilities, hence, they are bounded between $0$ and $1$. The details of this calculation are presented in Appendix  \ref{app:low_up_b}. The evaluation of the function on a lattice offers two advantages that are central for obtaining a feasible computation. First, the algorithm can be highly parallelized, as each lattice point can be evaluated independently of the others. Second, the algorithm can be made iterative: one first gets a rough estimate, in which many points of the lattice can already removed, and then refines the lattice for the remaining points. A representation of this method is shown in Fig.~\ref{fig:iter}. This allows us to estimate our functions on lattices of approximately $10^{18}$ and $10^{19}$ points. See Appendix~\ref{app:low_up_b} for further details.

A natural question arises regarding the relation between the saturation of the temporal inequality in Eq.~\eqref{eq:def_temp_ine} and the optimal accuracy of classical clocks. We see that the optimal clock in terms of accuracy, i.e., 
the one-way model, is also optimal for the finite sequence with $L=3$, but in the other cases it is outperformed by the cyclic model. The latter, in turn, has a non-optimal accuracy and, as shown in Sec.~\ref{sec:cont_lim}, does not have a continuous limit. Both models can be seen as a special instance of what we call the {\it multicyclic model} that generalizes to
arbitrary dimension, and for which $p(L)$, $\mu$, $\sigma$, and $R$ have analytic expressions in terms of the single parameter $q$. The multicylic model is discussed more in Sec.~\ref{sec:high_dim}, and its properties are summarized in Table~\ref{table:classical_models}. See also Appendix~\ref{app:class_mod} for more details on the derivation of such properties.

\subsection{Quantum violations}

\begin{figure}
\includegraphics[width=0.5\textwidth]{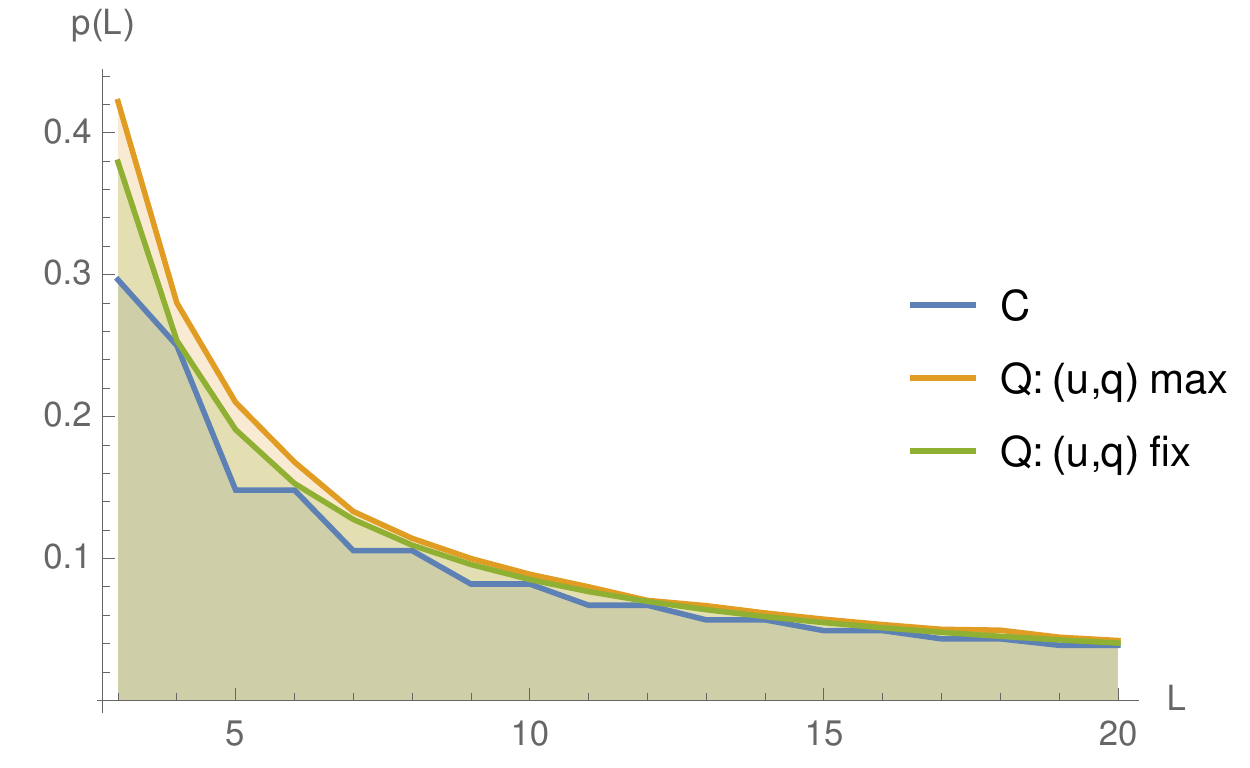}
\caption{\label{fig:plot_Q_C} Classical bound (blue) for $p(L)$ and quantum value for the model in Eq.~\eqref{eq:q_modelEU}, for $u=2 q/(1+q^2)$ and $q=1-2/L$ (green) or maximized over $u,q$ (yellow). }
\end{figure}

Again, given a classical bound as in Eq.~\eqref{eq:def_temp_ine} we can look for violations with a quantum model of the same dimension. For the qubit case, it turns out that the same model described in the previous section  is also able to violate the classical inequality in Eq.~\eqref{eq:def_temp_ine}. In fact, substituting in Eq.~\eqref{eq:q_modelEU} $u={2q}/( {1+q^2})$ and $q= \left( 1 - \frac 2 {\mu} \right)$, as before, and choosing $\mu=L$, one obtains a violation of the classical bound $p(L)\leq \Omega_{2,L}^C$. A maximization over $u$ and $q$ as free parameters, on the other hand, provides only a slightly better value. The result of this optimization are shown   in Fig.~\ref{fig:plot_Q_C}.  A similar model can be observed to outperform the multicyclic models also in higher dimensions. Details of the calculation are presented in Appendix~\ref{app:q_models}.

%%%%%%%%%%%%%%%%%%%%%%%%%%%%%%%%%%%%%%%%%%%%

%%%%%%%%%%%%%%%%%%%%%%%%%%%%%%%%%%%%%%%%%%%%

\section{Continuous limit of clock models}\label{sec:cont_lim}
We now investigate  the limit of continuous time  and how this recovers known results in the literature. 
Let us denote by $\delta>0$ the discrete time-step, and $\{\II_0^{(\delta)}, \II_1^{(\delta)}\}$ the corresponding quantum instrument. The continuous-time limit corresponds to $\delta\to 0^+$, so in order to characterize the case of ``no tick'' in a finite interval $[0,t]$, we need to consider a diverging number $N$ of application of the instrument, namely $N=t/\delta\in\nn$, with the composite instrument 
\begin{align}\label{eq:composite_defm}
\td \II_0^{(t)}:= \lim_{N\to +\infty} \left( \II_0^{(t/N)} \right)^{\circ N}.
\end{align}

To find the form of $\td \II_0^{(t)}$, we impose some very natural properties, which we can compactly write as
\begin{align}
\td \II_0^{(0)}=  {\rm id},\ (\text{ with }{\rm id}  \text{ the identity channel}),\label{eq:cond_cont_2}\\
\td \II_0^{(t_1+t_2)}=\td \II_0^{(t_1)} \circ \td \II_0^{(t_2)}  \quad \forall t_1, t_2 \geq 0,
\label{eq:cond_cont_3} \\
\lim_{t\to 0^+} \| \td \II_0^{(t)}-  {\rm id} \|=0, \label{eq:cond_cont_4}
\end{align} 
and interpret as follows. First, if no time has passed, nothing happens, Eq.~\eqref{eq:cond_cont_2}; second, the application is divisible, Eq.\eqref{eq:cond_cont_3}; third, we want the clock not to move instantaneously, Eq.~\eqref{eq:cond_cont_4}. For every map that satisfy these properties (see \cite{1982Gene}), there exists a liner operator $\hat L$ such that $\td \II_0^{(t)}$ forms a semigroup with a generator representation, namely $\td \II_0^{(t)} = \exp(t \hat L)$. It is then straightforward to verify that the choice $ \II_0^{(t/N)}= \exp\left((t/N) \hat L\right)$ in Eq.~\eqref{eq:composite_defm} implies that the conditions are satisfied. This guarantees self-consistency between definition~\eqref{eq:composite_defm} and the properties it must satisfy, i.e., \eqref{eq:cond_cont_2}-\eqref{eq:cond_cont_4}. At this stage, we have not used the fact that the map $\II_0^{(t/N)}$ is completely positive. Now, in order to prove that the generator $\hat L$ must be of the Lindblad form, it suffices to note that $\td \II_0^{(t)}$ is completely positive since $\II_0^{(t/N)}$ is. Therefore, by theorem 2.2 in \cite{CPGenerator}, it follows that the generator $\hat L$ must be of the Lindblad form, namely there exists Hamiltonian $H$, and operators $A_m$ such that for a $d$ dimensional clock
\begin{align}
\nonumber \mathcal{L}(\cdot)=-{i \over \hbar }[H,\cdot ]-{\frac {1}{2}}\left\{A_{0}^{\dagger }A_{0},\cdot \right\}+\sum _{n=1}^{d^2-1}A_{n} (\cdot) A_{n}^{\dagger }\\ -{\frac {1}{2}}\left\{A_{n}^{\dagger }A_{n},\cdot \right\},
\end{align}
with  $\td \II_0^{(t)}= \exp{t \mathcal{L}}$, and $\II_0^{(\delta)}= \exp{\delta \mathcal{L}}$. Note the addition of the operator $A_0$. This is because the channel is not trace preserving but rather trace non-increasing.  Therefore, the generator $\mathcal{L}$ can always be found by evaluating $ \lim_{\delta\to 0^+} ({ \II_0^{(\delta)} -{\rm id}})/{\delta}$.

The classical case is a special case of the above, and leads to the substochastic matrices $\td T_0^{(t)}= \me^{ t A_0}$, and $T_0^{(\delta)}= \me^{\delta A_0}$ where $A_0$ is a substochastic generator giving rise to an evolution $\pi(t)= \pi_1 \me^{t A_0}$ for $t\geq 0$.

From the continuity condition, $\lim_{\delta\to 0^+} \| T_0^{(\delta)}- {\rm id}\|=0$,  it is clear that the cyclic model does not have a continuous limit. In the one-way model, the condition is satisfied for $q(0)=1$. For $\delta\approx 0$, we can approximate $q$  via a Taylor expansion as $q(\delta)\approx 1- \alpha \delta$, with $\alpha > 0$. The condition on the first derivative comes from the fact that $q \leq 1$.\footnote{The first derivative equal to zero is excluded, since it is inconsistent with the assumption of a time-independent generator of the continuous model.}
 
For the $d$-dimensional cyclic model (see Appendix~\ref{app:class_mod}), the generator is  $ A_0 = \lim_{\delta\rightarrow 0^+} \frac{T_0 -\openone}{\delta} $ and consists of a bidiagonal matrix with $-1$ on the diagonal and $1$ on the upper diagonal. This is the \textit{ladder clock} defined in \cite{ATGRandomWalk} and proven to be the most accurate classical continuous-time clock in \cite{Woods2018} --- achieving an accuracy $R=d$. This clock can also be approached thermodynamically  \cite{ErkerPRX2017}, in the limit of semi-classical dynamics and  infinite entropy cost.

Similarly, we investigate the limit of the quantum model in Eq.~\eqref{eq:q_modelEU}, and compare it with the \emph{quasi-ideal clock} from \cite{Woods2021,WSO16,Woods2018}. For a single Kraus operator, a general continuous-time evolution can be written as $\ket{\psi(t)}=\me^{(iHt-Vt)}\ket{\psi(0)}$, with $H$ Hermitian 
and $V$ positive semidefinite. The discrete evolution is  
$\ket{\psi_n}=K_0^n \ket{\psi_0}$ with Kraus operator $K_0=U_0\sqrt{E_0}$ and we can put $K_0=\me^{\delta(-V+iH))}\approx \openone + \delta (-V+iH)$
for $\delta \approx 0$. From this relation and with
$q=q(\delta)$, we can associate, similarly to the classical case, $-V+iH = \lim_{\delta\rightarrow 0^+} \frac{K_0 -\openone}{\delta}$. By substituting explicitly 
the quantum model from Eq.~\eqref{eq:q_modelEU} and computing the limit (see 
Appendix~\ref{app:cont2}) we identify $V=(\openone+\sigma_z)/2$ and $H=\sigma_y/\sqrt 2$ and 
obtain in this way the continuous limit of our quantum model, which indeed closely resembles the quasi-ideal clock of~\cite{Woods2021,Woods2018,WSO16},
in that 
the Lindblad operator decomposes into a free evolution ($H$) and a measurement ($V$), 
with the corresponding basis of eigenvectors related by Fourier transform.

%%%%%%%%%%%%%%%%%%%%%%%%%%%%%%%%%%%%%%%%%%%%

%%%%%%%%%%%%%%%%%%%%%%%%%%%%%%%%%%%%%%%%%%%%

\section{Outlook on higher dimensional clocks}\label{sec:high_dim}

\begin{figure}[t]
	\includegraphics[width=0.5\textwidth]{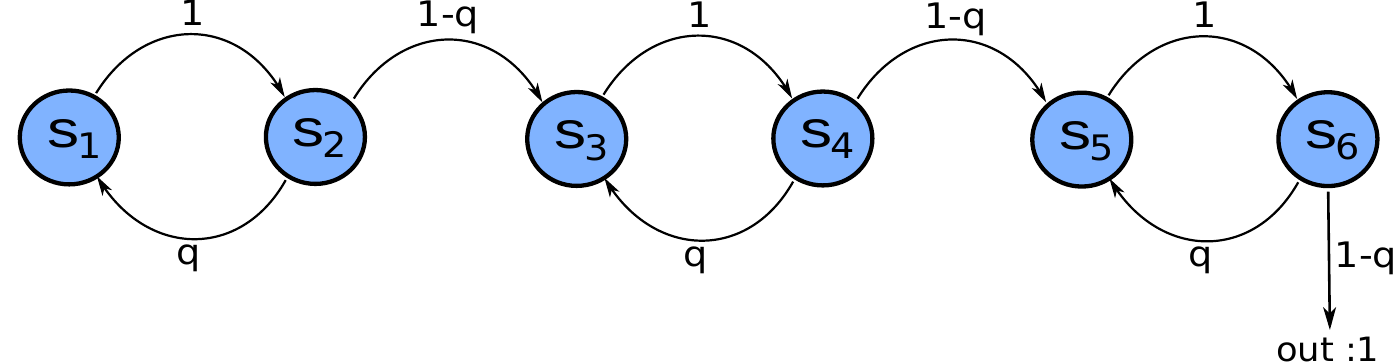}
	\caption{\label{fig:multi_cyc_App} Example of multicyclic model for $d=6$ and $k=2$. Within each block of length $k=2$, the machine cycles with probability $q$ or moves  forward with probability $1-q$. When the last state of the last block is reached, the machine can output $1$ with probability $1-q$.}
\end{figure}

The classical models arising for $d=2$ can be generalized to a family of higher dimensional models that we 
call {\it multicyclic models}. They are parametrized by a positive integer $k$, which 
gives the size of each block within which the behaviour is cyclic, but with the 
possibility of a transition from one block to the other, as in the one-way model. Only 
once the last state is reached, there is a nonzero probability of emitting the output 
``tick''. See Fig.~\ref{fig:multi_cyc_App} for a pictorial description and Appendix~\ref{app:class_mod} for further details. The one-way and cyclic models are recovered for $k=1$ and $k=d$, respectively. The main properties of the multicyclic model, namely, the expressions for the probability, mean, variance, and accuracy in terms of the model's parameters, are collected in Table~\ref{table:classical_models}.

\begin{table}[b]
\begin{center}
\begin{tabular}{ |c|c|c|c|c| } 
\hline
 model &  $\max p_{\pi_s}(L)\ $ & $\mu\ $ & $\sigma^2\ $ & $R\ $ \\
\hline
mc$_{d,k}$  & $\binom{m-1}{n-1} (1-n/m)^{m-n} (n/m)^n$ & $\frac{d}{1-q}$ & $\frac{k d q}{(1-q)^2}$ & $\frac{d}{k q}=\frac{d \mu}{k(d-\mu)}$ \\
\hline
\multicolumn{5}{|c|}{for $ n = \frac{d}{k},\ m=\ceil{\frac{L}{k}}^{\vphantom{L}}_{\vphantom{q}} ,\  s=(L\ {\rm mod}\ d) +1 \qquad$}  \\
\hline
\end{tabular}
\end{center}
  \caption{Summary of relevant quantities for the multicyclic model of dimension $d$ and block-size $k$ .}
  \label{table:classical_models}
 \end{table} 

We investigate the optimality of the multicyclic model via an extensive numerical search based on techniques from machine learning, specifically the Adam algorithm~\cite{Adam}. For Eq.~\eqref{eq:conj1} we search for a violation for  $d=3,\ldots,10$, and we explore the bounds for Eq.~\eqref{eq:def_temp_ine} for $d=3,\ldots,10$ and $L=d+1,\ldots,d+10$ and compare the results with the multicyclic model. In no case the numerical search provides better solutions. Details on numerical searches are collected in 
Appendix~\ref{app:num_search}. Moreover, the conjectured inequality presented in Eq.~\eqref{eq:conj1} is also supported by the optimality of the one-way model in the continuous-time limit \cite{Woods2018}.
These generalized classical models inspired quantum models beyond the qubit case, which are presented in Appendix~\ref{app:q_models}.
Recently, \cite{Vieira2021} extended the multicyclic model to the so-called {\it enhanced multicyclic model}, able to provide higher values for $p(L)$ for certain lengths and dimensions, specifically for $(L,d)=(7,5)$, $(9,7),(10,7)$. 
Moreover, the preprint also presents a high dimensional quantum model, based on the ideas described here, for the special case $L=d+1$ and saturating the algebraic maximum, $p(L)=1$, in the limit ${L\rightarrow \infty}$.

%%%%%%%%%%%%%%%%%%%%%%%%%%%%%%%%%%%%%%%%%%%%

%%%%%%%%%%%%%%%%%%%%%%%%%%%%%%%%%%%%%%%%%%%%

\section{Discussion and conclusions} 
We have investigated the operational notion of ticking clocks in the context of temporal sequences of measurements and have observed how the performance of such clocks is connected to a memory resource. We have shown that genuine quantum measurements  outperform classical ones as 
regularly-ticking devices by making a more efficient use of memory. It is natural to ask what is the relation between our results and present-day
time-keeping devices such as the atomic fountain clocks currently providing time and frequency standards~\cite{Heavner2014}. In simple terms, an atomic clock works by 
 tuning a microwave synthetizer to the frequency corresponding to the hyperfine atomic 
 transition of the ground state of the Caesium-133 atom. 
All relevant operations involve transitions or measurements in the energy eigenbasis, hence, they do not involve typical quantum elements such as non-commuting/incompatible 
measurements, and can straightforwardly be described by a classical hidden variable model. 
In this description, an atomic clock produces ticks, e.g., every second, by counting 
approximately $9 \times 10^9$ oscillations, for which a high
internal (classical) memory is needed. Our work instead, despite being still far from practical applications and involving simple (low memory) systems, focuses on {\it genuinely quantum} models, i.e., models that feature a more efficient generation of information exploiting quantum coherence, or, in other words, truly quantum temporal correlations. In this respect, our results complement recent literature~\cite{Woods2021,WSO16,Woods2018,ErkerPRX2017,SchwarzhansPRX2021,Pearson2020} in contributing to rethinking what are the fundamental quantum resources involved in the generation of 
time signals and possibly inspire future designs of truly quantum-improved clocks.

From the point of view of quantum clock models, our results confirm recent intuitions on optimal quantum clocks~\cite{Woods2018}, and extend them to the discrete-time scenario,  revealing its richer mathematical structure: We found that  new optimal classical models arise that do not have a 
continuous-time limit, a feature  absent in the quantum case, where the
continuous limit is always valid. This suggests that some fundamental
difference between classical and quantum models arises also in this
respect.

 Complementing recent works~\cite{Woods2018,ErkerPRX2017,SchwarzhansPRX2021}, we understand quantum advantages in a general framework of nonclassical temporal correlations.
 Notwithstanding the central role of LGI in a foundational perspective, such inequalities are not suitable for discussing tasks involving microscopic systems and invasive operations. By allowing for a relaxation of NIM to  ``limited invasivity'', we are able to discuss nonclassical temporal correlations in relation with technological applications such as the design of time-keeping devices. 

Interestingly, the gap found between classical and quantum clocks also reveals what is the {\it simplest, yet nontrivial} temporal correlation scenario, namely, a scenario with just $1$ input and $2$ outputs. This is in contrast with standard Bell and Leggett-Garg inequalities, where at least $2$ inputs are needed to find a gap between classical and quantum correlations. 
This is due to the nontrivial causal structure associated with the finite-state machine model (not even representable by a directed acyclic graph), which is analogous to recently investigated nontrivial causal structures in nonlocality where no inputs are needed to witness a gap between classical and quantum correlations (e.g., the triangle scenario analysed in~\cite{RenouPRL2019}). 
The analytical and numerical tools developed here, together with the limited number of parameters of our models, suggest that it may be possible to derive inequalities for arbitrary sequences and dimensions. This will be the object of future research.

Finally, finite and discrete-time sequences are, similarly to Bell inequalities, more experimentally friendly and we hope our results will stimulate also an experimental investigation of temporal correlations in the finite memory scenario in the near future.

\acknowledgments We thank Edgar Aguilar, Niklas Johansson, Matthias Kleinmann, Jan-{\AA}ke Larsson, Miguel Navascu\'es, Nikolai Miklin, and Lucas Vieira for fruitful discussions. In particular, we thank Edgar Aguilar for finding the cyclic model and showing that it outperforms the one-way model for the $p(L)$ expression. CB and GV have been supported by the Austrian Science Fund (FWF) through projects ZK 3 (Zukunftskolleg) F7113 (BeyondC) and M 2462-N27 (Lise-Meitner). MW acknowledges support from the Swiss National Science Foundation (SNSF) via an AMBIZIONE Fellowship (PZ00P2\_179914) in addition to the National Centre of Competence in Research ``QSIT''.

%%%%%%%%%%%%%%%%%%%%%%%%%%%%%%%%%%%%%%%%%%%%%%%%%%%%%%%%%%%%%%%%%%%%%%%%%

%%%%%%%%%%%%%%%%%%%%%% APPENDIX %%%%%%%%%%%%%%%%%%%%%%%%%%%%%%%%%%%%%%%%%

%%%%%%%%%%%%%%%%%%%%%%%%%%%%%%%%%%%%%%%%%%%%%%%%%%%%%%%%%%%%%%%%%%%%%%%%%
\appendix

\input{Appendix.tex}

\bibliography{TIC}{}

\end{document}

%% file: MischaCommandsV2.tex
\usepackage{dsfont}
\usepackage{bbm} % WARNING: JNP has problems with this package when submitting PDFs generated using it. i.e. using the \id command
\usepackage[normalem]{ulem} % For using the \sout{Hellow word} command to strike-out text.
\usepackage{color} % for \color commands
\newcommand{\be}{\begin{equation}}
\newcommand{\ee}{\end{equation}}
\newcommand{\nn}{{\mathbbm{N}}}

\newcommand{\me}{\mathrm{e}}

\def\ba#1\ea{\begin{align}#1\end{align}} % use "\ba" and "\ea to begin/end equation environment "align"

\newcommand\mpwS[1]{MW:{\let\helpcmd\sout\parhelp#1\par\relax\relax} }
\long\def\parhelp#1\par#2\relax{%
	\helpcmd{#1}\ifx\relax#2\else\par\parhelp#2\relax\fi%
}
\newcommand{\nocontentsline}[3]{}
\newcommand{\tocless}[2]{\bgroup\let\addcontentsline=\nocontentsline#1{#2}\egroup}

%% file: Appendix.tex
\section{Derivation of optimal bit clock}\label{app:2_dim_cl}

Let us consider 2-state classical machine with an arbitrary transition matrix 
$T_0=\left(
\begin{array}{cc}
a & b \\
c & d 
\end{array}
\right) .
$
Computing the inverse of $\id-T_0$ and fixing the initial state to be $\pi_0=(1,0)$ we can directly compute the accuracy as a function of the model parameters $a,b,c,d$ via Eq.~\eqref{eq:muqzcl}. We obtain
\begin{widetext}
\begin{equation}\label{eq:mean_var_ow}
\mu = \frac{-b+d-1}{-ad +a+b c+d-1},\qquad \sigma^2=\frac{a \left((d-1)^2-b (d+1)\right)+b (b (c-1)-c d+3 c+d+1)}{(-ad+a+b c+d-1)^2},
\end{equation}
and in turn the accuracy reads
\begin{equation}
R= \frac{\mu^2}{\sigma^2}= \frac{(b-d+1)^2}{a \left((d-1)^2-b (d+1)\right)+b (b (c-1)-c d+3 c+d+1)}.
\end{equation}
The above expression can be optimized analytically for any fixed value of $\mu > 2$, while for $\mu \leq 2$ the accuracy tends to infinity. %Moreover, it is intuitive that an optimization of $R$ with $\mu$ as a free parameter will return a minimal value of $\mu$. This intuition is confirmed by the exact solution discussed below. 
To perform the optimization with a fixed mean $\mu$, it is helpful to express the parameter $d$ as a function of $a,b,c,\mu$:
\begin{equation}
d= \frac{a \mu+b c \mu+b-(\mu-1)}{a \mu-(\mu-1)},
\end{equation}
which can be verified by direct substitution in Eq.~\eqref{eq:mean_var_ow}. 

With the above substitution, we are now able to prove that the solution is given by the one way model, namely  $b=2/\mu$, $a=d=1-b$, $c=0$, where our parameter $q$ appearing in Eq.~\eqref{eq:Ropt} of the main text is simply $q=a=1-2/\mu$.
The proof is based on the Karush-Kuhn-Tucker (KKT) conditions \cite{CVXBook}, a generalization of Lagrange multiplier for inequality constraints, that can be applied to a problem of the form
\begin{equation}
\begin{split}
\text{Maximize: }& f(x) \\
\text{subject to: }& g_i(x) \leq 0,\ i=1,\ldots,m \ ,
\end{split}
\end{equation}
where the optimization is performed over a convex set in $\mathbb{R}^n$ and both the objective function $f$ and the constraints $g_i$ satisfy some regularity conditions. In our case, the constraints $g_i$ are all affine functions of the parameters, which guarantees that for a global maximum $x^*$ the KKT conditions are satisfied~\cite{CVXBook}.

A necessary condition for a point $x^*$ to be a global maximum is that there exists $\lambda_i$ for $i=1,\ldots,m$ such that
\begin{equation}
\begin{split}
\nabla f(x^*) - \sum_{i=1}^m \lambda_i \nabla g_i(x^*)  = 0,\\
g_i(x^*)\leq 0, \text{ for all } i,\\
\lambda_i \geq 0, \text{ for all }i,\\
\lambda_i g_i(x^*) = 0, \text{ for all }i.
\end{split}
\end{equation}
In our case, we want to optimize the function $R(a,b,c)$ for a given $\mu$, namely
\begin{equation}
R(a,b,c)=\frac{b \mu^2 (-a+c+1)}{\mu ((a-1) \mu (2 a+b-2)+a b+4 a+b c \mu-b c+b-4)+2}
\end{equation}
with the constraints $g_i(a,b,c)\leq 0$ defined via
\begin{equation}
g_1=-a,\ g_2=-b,\ g_3 = -c,\ g_4= a+b-1, \ g_5=c-1 .
\end{equation}
Now the set of equations
\begin{equation}
\begin{split}
\nabla R(x) - \sum_{i=1}^5 \lambda_i \nabla g_i(x) = 0,\\
g_i(x)\leq 0, \ i=1,\ldots,5,\\
\lambda_i \geq 0,\ i= 1,\ldots,5, \\
\lambda_i g_i(x) = 0, \ i =1,\ldots,5
\end{split}
\end{equation}
can be solved explicitly, with the aid of a computer algebra system,  for $\mu > 2$ and give, for the case $b=0$, i.e., $g_2=0$, the solution
\begin{equation}
 \mu > 2,\  a=1, \ b= c= \lambda_1= \lambda_2 = \lambda_3=  \lambda_4=  \lambda_5=0 ,
\end{equation}
and for the case $b>0$, the solution
\begin{equation}
\begin{split}
&\mu > 2,\ b=\frac{2}{\mu}, \ a=1-b,\  c=\lambda_1 = \lambda_2=\lambda_5=0, \ \lambda_3= \lambda_4 =\frac{2 b\mu^2 (b\mu-1)^2}{(b\mu (b (m-1)-2)+2)^2}.
\end{split}
\end{equation}
\end{widetext}
It is straightforward to check that only the second solution corresponds to a maximum, giving 
$R=\frac{2\mu}{\mu -2 }$, and it is precisely the one obtained by the one-way model in the bit case.

Finally, it is interesting to compute explicitly the probability $p(L)$, as a function of $L$ and the model parameter $a,b,c,d$ for the bit case. This amounts to calculating explicitly the integral form
\begin{equation}\label{eq:pLfromZ}
p(L)= \pi_0 \left(\frac 1 {2\pi i} \oint_C z^{L-1}(1-z)  \left( z\openone - T_0 \right)^{-1}  {\rm d} z \right) \eta ,
\end{equation}
from the residues of the function 
\begin{equation}\label{eq:res_gen}
\begin{split}
z^{L-2}(1-z)\tilde f_{\rm cl}(z) &=z^{L-1}(1-z) \pi_0  \left( z\openone -  T_0 \right)^{-1} \eta \\
&= \frac{z^{L-1}(1-z)}{\det(z\openone -  T_0)} \pi_0   {\rm adj}(z\openone - T_0)   \eta ,
\end{split}
\end{equation}
where ${\rm adj}(A)$ denotes the adjugate matrix of $A$, i.e., $[{\rm adj}(A)]_{ij} = (-1)^{i+j} M_{ji}$, where $M_{ji}$ is the $(j,i)$-minor of the matrix $A$, namely, the determinant of the matrix obtained by deleting the row $j$ and column $i$ of $A$ (see, e.g., \cite{Horn2012matrix} for a textbook reference). In the bit case,  the adjugate of $z\openone - T_0$ is 
\begin{equation}
{\rm adj}(z\openone - T_0)=\left(
\begin{array}{cc}
z-d & b \\
c & z-a 
\end{array}
\right) ,
\end{equation}
and thus, fixing the initial state as $\pi_0=(1,0)$, we have $\pi_0 \cdot {\rm adj}(z\openone - T_0) \cdot \eta=(z-d+b)$.
By Eq.~\eqref{eq:res_gen}, we have
\begin{equation}\label{eq:sumresidues}
z^{L-2}(1-z)\tilde f_{\rm cl}(z)=\frac{z^{L-1}(1-z)}{(z-t_+)(z-t_-)} (z-d+b) ,
\end{equation}
where $t_\pm = \tfrac 1 2 \left(a+d \pm \sqrt{\Delta}\right)$ with $\Delta=(a-d)^2+4bc$ are the eigenvalues of $T_0$. 

First, let us assume that $T_0$ has two distinct eigenvalues $t_+\neq t_-$.
Calculating the residues of the function in Eq.~(\ref{eq:sumresidues}) we obtain the expression for the probability as
\begin{equation}\label{eq:pL_general}
\begin{aligned}
p(L)&=\frac{t_+^{L-1}(1-t_+)}{t_+-t_-} (t_+-d+b)-\frac{t_-^{L-1}(1-t_-)}{t_+-t_-} \\
&\times (t_--d+b) \\
&=\frac{1}{2^{L+1}\sqrt{\Delta}}\left[(\sqrt{\Delta}-a-d+2)(\sqrt{\Delta}-a-2 b+d) \right.\\
&\times (-\sqrt{\Delta}+a+d)^{L-1} -(\sqrt{\Delta}+a+2 b-d )\\
&\left. \times   (\sqrt{\Delta}+a+d-2)  (\sqrt{\Delta}+a+d )^{L-1} \right] .
\end{aligned}
\end{equation}

In the case of identical eigenvalues, i.e., $t_+=t_-=t_0$ either $T_0$ is proportional to the identity, i.e., $a=d=t_0$ and $b=c=0$, or $T_0$ is not diagonalizable, i.e., $a=d$ and $c=0\neq b$. The former case is trivial and the latter can be computed similarly to the previous case. Notice, however, that the   function $z^{L-1}(1-z) (z-d+b)/(z-t_0)^2$ has a second order pole in $t_0$, so we cannot directly substitute $t_+,t_-\mapsto t_0$ in Eq.~\eqref{eq:pL_general}, but the residue must be computed with the formula for the second order poles.

\section{Classical clock models for general dimension}\label{app:class_mod}
In the following, we  discuss in detail the classical models presented in the main text, which are all special instances of what we call the {\it multicyclic model}.
As we discussed in the main text, in the bit case two classical models arise, namely, the one-way model and the cyclic model. It is instructive to summarize their main properties to understand how to generalize them. In the one way model, the machine either remains in the same state with probability $q$ or transitions to the subsequent state with probability $1-q$, always emitting the output $0$. When the last state is reached, the output $1$ is emitted with probability $1-q$. In the cyclic model, the machine transitions from one state to the next with probability one always emitting the output $0$, except in the last state where it may cycle, i.e., go back to the first state emitting $0$, with probability $q$, or emit the output $1$ with probability $1-q$.

The multicyclic model generalizes both ideas to arbitrary dimension. A simple example for dimension $6$ is depicted in Fig.~\ref{fig:multi_cyc_App} and is described by the matrix:
\begin{equation}\label{eq:multicyc6}
T_0^{{\rm mc},k} =
\left(
\begin{array}{cc|cc|cc}
 0 & 1 & 0 & 0 & 0 & 0 \\
 q & 0 & 1-q & 0 & 0 & 0 \\
 \hline
 0 & 0 & 0 & 1 & 0 & 0 \\
 0 & 0 & q & 0 & 1-q & 0 \\
 \hline
 0 & 0 & 0 & 0 & 0 & 1 \\
 0 & 0 & 0 & 0 & q & 0 \\
\end{array}
\right).
\end{equation}

The behaviour of the machine can be interpreted as follows. 

(i) The matrix consists of blocks of size $k$ ($k=2$ in the example of Eq.~\eqref{eq:multicyc6} and Fig.~\eqref{fig:multi_cyc_App}). Starting from the first state, $\pi_1:=(1,0,\ldots,0)$, the machine transitions to the next state with probability $1$, always emitting the output $0$. 

(ii) When the last state of the first block is reached, the machine may cycle with probability $q$, i.e., go back to the first state of the block, or transition to the next block with probability $1-q$, again emitting the output $0$ with probability $1$. The same behaviour is repeated in the second block, and all the other blocks except the last one.

(iii) When the last state of the last block is reached, the machine may cycle to the first state of the block with probability $q$ and emit the output $0$, or emit the output $1$, for which the sequence is terminated, so the subsequent state needs not to be specified.

For the bit case, i.e., $d=2$, the multicyclic model corresponds to the one-way model in the case of blocks of size $k=1$, and to the cyclic model in the case of one block of size $k=2$. We generalize this terminology to arbitrary dimensions, i.e., we call a multicyclic model with $k=1$ a {\it one-way} model and one with $k=d$ a {\it cyclic} model. 

\subsection{Probability $p(L)$}

To compute the probability $p(L)$, let us first consider the case in which $d = n k$ and $L= m k$ with $k,m,n\in \mathbb{N}^+$, and $m>n$. Then, the probability $p(L)$ can be written as
\begin{equation}\label{eq:pL_multic}
p_{\rm mc}(L)= \pi_1 T_0^L (\openone - T_0) \eta=\binom{m-1}{n-1} (1-q)^n q^{m-n}, 
\end{equation}
valid for  $m=\frac{Ln}{d}=\frac{L}{k}$.The expression in Eq.~\eqref{eq:pL_multic} can be understood as follows. The output $1$ is generated only in the last state, giving a factor $(1-q)$ to the total expression. A factor $(1-q)^{n-1}$ comes from the probability of transitioning from the first to the last block, as each transition contributes to a factor $(1-q)$ and we need $n-1$ of them, $n$ being the number of blocks. The binomial coefficient $\binom{m-1}{n-1}$ comes as a combinatorial factor from the possible choices of $n-1$ transitions out of $m-1$ possibilities, as the total length is $L=m k$ and each block has size $k$. Finally, $q^{m-n}$ is the probability of cycling $m-n$ times, i.e., in the remaining $(m-n)k$ steps, in order to output $1$ at the correct step $L=m k$. 

The expression in Eq.~\eqref{eq:pL_multic} can be optimized over $q$ simply by taking the derivative of the expression $(1-q)^n q^{m-n}$, giving
\begin{equation}
q_{\rm opt} = 1-\frac{n}{m}=1-\frac{d}{L}.
\end{equation}

For a fixed initial state $\pi_1=(1,0,\ldots,0)$,  the length of the sequence must be a multiple of the block length $k$ in order to have a non-zero probability of outputting $1$ at the correct time, namely
\begin{equation}\label{eq:prob01_mc}
p_{{\rm mc}, \pi_1}(L) =\left\lbrace
\begin{array}{cc}
\binom{m-1}{n-1} (1-q)^n q^{m-n} & \text{ if } \frac{L}{k}=m\in \mathbb{N}^+,\\
0 & \text{ otherwise }
\end{array}
\right. .
\end{equation}
In order to maximize $p(L)$ for different lengths, one may decide to start from a different initial state, within the first block, namely, from the  $s$-th state 
$\pi_s$ where $s + L -1 = 0\ {\rm mod}\ k$, and the $-1$ comes from starting counting from $1$, i.e, 1st, 2nd, etc. Intuitively, since the first $k-1$ transitions are deterministic, the probability obtained with the initial state 
$\pi_s$ is equivalent to the probability obtained starting from the initial state $\pi_1$ for length $L+s-1$. As a consequence, instead of  $m=\frac{L}{k}$ appearing in Eq.~\eqref{eq:pL_multic}, we have a factor $ m=\ceil{\frac{L}{k}}$. Alternatively, one can verify by direct computation that for each block of size $k$
\begin{equation}\label{eq:t0c}
B_k:=\left(
\begin{array}{ccccc}
0 & 1 &0 &\ldots & 0\\
0 & 0 & 1 & \ldots &0\\
\vdots & & \ddots & \ddots & \vdots\\
0& \ldots & & 0 & 1 \\
q& \ldots & & 0 & 0
\end{array}
\right), \ 
\end{equation}

where the $n$-th power is given by 
\begin{equation}
[B_k^n]_{ij} =\left\lbrace
\begin{array}{cc}
q^{\floor{\frac{n+i-1}{k}}} & \text{ if } j = (n+i-1 \mod k) + 1,\\
0 & \text{ otherwise. }
\end{array}
\right.
\end{equation}

We thus have the general expression for the multicyclic model of dimension $d$ and block size $k$, with $\frac{d}{k}=n\in \mathbb{N}^+$ 
\begin{equation}
\begin{split}
p_{{\rm mc},\pi_s}(L) &= \binom{m-1}{n-1} (1-q)^n q^{m-n}\\
&=\binom{m-1}{n-1} \left(\frac{n}{m}\right)^n \left(1-\frac{n}{m}\right)^{m-n}\\
\text{  for } m &=\ceil{\frac{L}{k}},\ s=(L\ {\rm mod}\ d) +1 .
\end{split}
\end{equation}
As an example, we provide in Table~\ref{tab:opt_dim} the optimal $k$ maximizing the expression $p(L)$ for $d=3,\ldots,10$ and ${L=d+1,\ldots,d+10}$, obtained simply by comparing all multicyclic models with $k n = d$. 
\begin{table*}[t]
\begin{tabularx}{\textwidth}{| X | X | X | X | X | X | X | X | X | X | X |}
\hline
\multicolumn{11}{|c|}{Optimal block-size $k$ for given $d$ and $L$}\\
\hline
\diagbox[width=3em]{$\ d$}{\vphantom{$\frac{L}{L}$} $L \ $} & d+1 & d+2 & d+3 & d+4 & d+5 & d+6 & d+7 & d+8 & d+9 & d+10\\
\hline
3 & 1 & 3 & 3 & 3 & 3 & 3 & 3 & 3 & 3 & 3\\
\hline
4 & 1 & 2 & 4 & 4 & 4 & 4 & 4 & 4 & 4 & 4\\
\hline
5 & 1 & 5 & 5 & 5 & 5 & 5 & 5 & 5 & 5 & 5\\
\hline
6 & 1 & 2 & 3 & 6 & 6 & 6 & 6 & 6 & 6 & 6\\
\hline
7 & 1 & 7 & 7 & 7 & 7 & 7 & 7 & 7 & 7 & 7\\
\hline
8 & 1 & 2 & 4 & 4 & 8 & 8 & 8 & 8 & 8 & 8\\
\hline
9 & 1 & 3 & 3 & 9 & 9 & 9 & 9 & 9 & 9 & 9\\
\hline
10 & 1 & 2 & 5 & 5 & 5 & 10 & 10 & 10 & 10 & 10\\
\hline
\end{tabularx}
\caption{\label{tab:opt_dim}Optimal block-size $k$ maximizing $p(L)$ for the multicyclic model for dimension $d=3,\ldots,10$ and length $L=d+1,\ldots,d+10$. The case $k=1$ and $k=d$ correspond, respectively, to the one-way and the cyclic model. The other cases, e.g., $(d,L) = (4,6), (6,8), (6,9),\ldots $, correspond to the multicyclic model with blocks of size $k\neq 1,d$. Typically, the one-way model ($k=1$) is optimal for $L=d+1$, nontrivial block sizes, i.e., $k\neq 1,d$ appear when $d$ and $L$ have some common factor, e.g., $d=4,6,8$, but not necessarily, e.g., $d=10$, in all other cases, the cyclic model is optimal ($k=d$). }
\end{table*}

\begin{widetext}
\subsection{Accuracy}

Using Eq.~\eqref{eq:prob01_mc}, we can compute directly the mean and variance of the distribution $p(L)$ and consequently its accuracy $R$, for a multicyclic model of dimension $d$ and block size $k$, with $d=n k$, $n,k\in \mathbb{N}^+$. First, let us notice that from the normalization condition
\begin{equation}
1=\sum_{L=1}^\infty p(L)=\sum_{L=d}^\infty p(L) =\sum_{m=n}^\infty \binom{m-1}{n-1} q^{m-n} (1-q)^n = \frac{(1-q)^n}{q^{n-1}}\sum_{m=n}^\infty \binom{m-1}{n-1} q^{m-1},
\end{equation}
we obtain
\begin{equation}\label{eq:norm_id}
\sum_{N = k}^\infty \binom{N}{k}y^{N}=\frac{y^k}{(1-y)^{k+1}}, \text{ for all } y\in [0,1).
\end{equation}
We can now proceed to calculate $\mu$.
\begin{equation}
\begin{aligned}\label{eq:mu_mc}
\mu&=\sum_{m=n}^\infty m k \binom{m-1}{n-1} q^{m-n} (1-q)^n= \frac{k n (1-q)^n}{q^n}\sum_{m=1}^\infty \frac{m}{n}\binom{m-1}{n-1} q^{m} \\
&=  \frac{ k n (1-q)^n}{q^n}\sum_{m=n}^\infty \binom{m}{n} q^{m}= \frac{k n (1-q)^n}{q^n} \frac{q^n}{(1-q)^{n+1}}
=\frac{k n}{(1-q)} ,
\end{aligned}
\end{equation}
where we used the identity in Eq.~\eqref{eq:norm_id}.
Similarly, we can compute the variance and obtain
\begin{equation}
\sigma^2 = \sum_{L=1}^\infty \left(L-\mu\right)^2 p(L)=k^2\sum_{m=n}^\infty \left(m-\frac{n}{(1-q)}\right)^2 \binom{m-1}{n-1} q^{m-n} (1-q)^n.
\end{equation}
By using the following identity 
\begin{equation}
\begin{aligned}
\sum_{m=n}^\infty m(m+1) \binom{m-1}{n-1} q^{m-n} (1-q)^n &= \frac{(1-q)^n n(n+1)}{q^{n+1}} \sum_{m=n}^\infty \binom{m+1}{n+1} q^{m+1}\\
&= \frac{(1-q)^n n(n+1)}{q^{n+1}} \frac{q^{n+1}}{(1-q)^{n+2}}
=\frac{n(n+1)}{(1-q)^2},
\end{aligned}
\end{equation}
we can write
\begin{equation}\label{eq:sig_mc}
\begin{aligned}
\sigma^2&=\sum_{L=d}^\infty \left(L-\frac{k n}{(1-q)}\right)^2 p(L) = 
k^2\sum_{m=n}^\infty \left[m(m+1) + m\left(\frac{2n}{(1-q)}+1\right) + \frac{n^2}{(1-q)^2}\right] \binom{m-1}{n-1} q^{m-n} (1-q)^n \\
&= k^2\ \frac{n(n+1)-n(2n+(1-q))+n^2}{(1-q)^2}= \frac{k^2n q}{(1-q)^2} .
\end{aligned}
\end{equation}
The accuracy can, then, be written as
\begin{equation}\label{eq:acc_mc}
R=\frac{\mu^2}{\sigma^2}=\frac{n}{q}=\frac{d}{k q}.
\end{equation}
We then recover the fact that the accuracy is optimal for the one-way model, i.e., $k=1$, as it is for the bit case, whereas the cyclic model gives the worst accuracy, i.e., $R=1/q$, which is even independent of the dimension.

\end{widetext}
\section{Quantum models}\label{app:q_models}

Here we present an explicit construction of quantum clock models that are able to 
outperform the optimal classical clock in $d=2$ and the multicyclic clocks in $d=3$. 
First, let us derive in detail the qubit clock model described in the main text
and observe how it violates the classical bounds on our temporal inequalities in $d=2$. 

\subsection{Qubit clock model}\label{sec:qubit_case}
An explicit quantum model that violates the inequality (7) of the main text for arbitrary values of $\mu$ is constructed as follows: consider a qubit in the initial state $\ket{\psi}=(1,0)$ that evolves via a single Kraus operator $K_0$ associated with the outcome $0$, i.e., $\II_0(\rho)=K_0 \rho K_0^\dagger$, defined as $K_0 = U_0 \sqrt{E_0}$, with 
\begin{equation}
E_0:=\left(
\begin{array}{cc}
1 & 0 \\
0 & q^2 
\end{array}
\right), 
\qquad 
U_0:=\left(
\begin{array}{cc}
\sqrt{u} & \sqrt{1-u} \\
 -\sqrt{1-u} & \sqrt{u} \\
\end{array}
\right).
\end{equation}
The mean value and variance of the corresponding distribution can be obtained via the $Z$-transform method, where now the moment generating function is
\begin{equation}
Q(z) = -(1-1/z) \mathcal Z[f(L)]:= -(1-1/z) \tilde f_{\rm qbit}(z) ,
\end{equation}
where 
\begin{equation}
f_{\rm qbit}(L)=\bra{\psi} (K_0^\dagger)^L K_0^L \ket{\psi} .
\end{equation}
To write down the above expression, it is helpful to consider $K_0$ in its Jordan normal form $K_0=P \Lambda_K P^{-1}$, where for our model we have $\Lambda_K = {\rm diag}(\kappa_+ , \kappa_-)$, i.e., the Kraus operator is diagonalizable. 
In particular, we have that $K_0$ has eigenvalues $\kappa_{\pm}=\tfrac 1 2 \left[(1+q)\sqrt{u}\pm \sqrt{\Gamma_{q,u}}\right]$, where $\Gamma_{q,u}=(1+q)^2u-4q$ and (not orthonormal) eigenvectors given by $\left(v_\pm, 1 \right)$ where
\begin{equation}
\begin{aligned}
v_\pm &= \frac{-(1-q)\sqrt{u} \pm  \sqrt{\Gamma_{q,u}}}{2\sqrt{1-u}} ,
\end{aligned}
\end{equation}
which corresponds to the columns of $P$.
Given this decomposition we write $(K_0^\dagger)^L K_0^L = (P^{-1})^T \Lambda_K^L P^T P \Lambda_K^L P^{-1}$, where
we used the fact that our $P$ is real, and we define $\ket{P^{-1}\psi}= P^{-1}\ket{\psi} = \left(1/(v_+-v_-), -1/(v_+-v_-) \right)=\left(\frac{\sqrt{1-u}}{\sqrt{\Gamma_{q,u}}}, -\frac{\sqrt{1-u}}{\sqrt{\Gamma_{q,u}}} \right)$.
This way we have
$f_{\rm qbit}(L)=\bra{P^{-1}\psi} F_L \ket{P^{-1}\psi}$ with 
 \begin{equation}
 \begin{split}
F_L:&= \Lambda_K^L P^T P \Lambda_K^L\\
&=\left(
\begin{matrix}
(1+v_+^2)\kappa_{+}^{2L} & (1+v_+ v_-)\kappa_+^L \kappa_-^L \\
(1+v_+ v_-)\kappa_+^L \kappa_-^L & (1+v_-^2)\kappa_{-}^{2L}
\end{matrix}
\right) \\
&= \left(
\begin{matrix}
(1+v_+^2)\kappa_{+}^{2L} & (1+q)q^L \\
(1+q)q^L & (1+v_-^2)\kappa_{-}^{2L} 
\end{matrix}
\right) ,
\end{split}
\end{equation}
where we used that $v_+ v_- =\kappa_+ \kappa_-=q$.
At this point, we can calculate the probability as 
\begin{equation}
p(L) = f_{\rm qbit}(L-1)-f_{\rm qbit}(L) = \bra{P^{-1}\psi} (F_L-F_{L-1}) \ket{P^{-1}\psi},
\end{equation}  
which results in
\begin{equation}
\begin{split}
p(L) = \frac{1-u}{q^2 \Gamma_{q,u}} ((1+v_-^2)\kappa_{-}^{2L}(\kappa_{+}^2-q^2)-2(1-q^2)q^{L+1}\\
+(1+v_+^2)\kappa_{+}^{2L}(\kappa_{-}^2-q^2)) .
\end{split}
\end{equation}
Furthermore, we can now calculate $\tilde f_{\rm qbit}(z)$ from the element-wise $Z$-transform of 
$F_L$, which is given by
  \begin{equation}
\tilde F(z)=\left(
\begin{matrix}
(1+v_+^2)\frac{1}{1-\kappa_{+}^2 z^{-1}} & (1+q)\frac{1}{1-q z^{-1}} \\
(1+q)\frac{1}{1-q z^{-1}} & (1+v_-^2)\frac{1}{1-\kappa_{-}^2 z^{-1}}
\end{matrix}
\right), 
\end{equation}
where we used the known $Z$-transform $\mathcal Z[a^L]=\frac{1}{1-az^{-1}}$.

Now, for calculating the mean and variance of our probability distribution 
we need to calculate $\tilde f_{\rm qbit}(1)$ and $\tilde f_{\rm qbit}^\prime (1)$. 
For the former, we can just substitute the value $1$ into $\tilde F(z)$ and multiply by the vectors $\bra{P^{-1}\psi}$ and $\ket{P^{-1}\psi}$, and we obtain
\begin{equation}
\begin{aligned}
\tilde f_{\rm qbit}(1) &=
\frac{1-u}{\Gamma_{q,u}}\left( 2\frac{q+1}{q-1}-\frac{\left(v_+^2+1\right)}{\left(\kappa_+^2-1\right)}-\frac{\left(v_-^2+1\right)}{\left(\kappa_-^2-1\right)}\right) =\\
&=\frac{(q-2) q u-(u-2)}{\left(q^2-1\right) (u-1)} .
\end{aligned}
\end{equation}
Similarly, for calculating the first derivative we can first derivate $\tilde F(z)$ entrywise:
  \begin{equation}
\tilde F^\prime(z)=\left(
\begin{matrix}
-\frac{(1+v_+^2)\kappa_{+}^2}{(\kappa_{+} -z)^2} & -\frac{q(1+q)}{(z-q)^2} \\
-\frac{q(1+q)}{(z-q)^2} & -\frac{(1+v_-^2)\kappa_{-}^2}{(\kappa_{-} -z)^2}
\end{matrix}
\right), 
\end{equation}
and then substitute the value $1$ and multiply, obtaining
\begin{widetext}
\begin{equation}\label{eq:fqbitprime1}
\begin{aligned}
f_{\rm qbit}^\prime (1) &= \frac{1-u}{\Gamma_{q,u}}\left(2\frac{q(q+1)}{(q-1)^2}
-\frac{\left(v_+^2+1\right)\kappa_+^2}{\left(\kappa_+^2-1\right)^2} 
-\frac{\left(v_-^2+1\right)\kappa_-^2}{\left(\kappa_-^2-1\right)^2}\right) \\
&=-\frac{q \left\lbrace 2 (q+1) u^2+qu [q(q-4)+1]-6u+3q\right\rbrace+1}{\left(q^2-1\right)^2 (u-1)^2} .
\end{aligned}
\end{equation}
In turn, those lead to the following expressions for the mean and variance:
\begin{align}
\mu &= \tilde f_{\rm qbit}(1) =\frac{q u(q-2) - (u-2)}{\left(q^2-1\right) (u-1)},  \label{eq:muqzqubit} \\
\sigma^2  &=-\mu(1-\mu)- 2\tilde f_{\rm qbit}^\prime (1) = \frac{2 \left(q^2+1\right) q u^2+[q(q-6)+1] \left(q^2+1\right) u+4 q^2}{\left(q^2-1\right)^2 (u-1)^2} .  \label{eq:sigmaqubit}
\end{align}
The expression on the left hand side of Eq.~\eqref{eq:ti_mu} of the main text reads
\begin{equation}\label{eq:expLHSqubit}
\mu (\mu-2) -2\sigma^2 = 3\mu^2-4\mu +4\tilde f_{\rm qbit}^\prime (1) = \frac{-4 q^2-(q \{q [q (q+4)-6]+4\}+1) u^2+8 [(q-1) q+1] q u}{\left(q^2-1\right)^2 (u-1)^2}  .
\end{equation}
and we can try to maximize it over the parameters $q$ and $u$ for each fixed value of $\mu$.
Note that, for fixed $\mu$ the above expression can be maximized by just minimizing $\tilde f_{\rm qbit}^\prime (1)$. 
At first, since we want $\mu$ to be fixed, we can invert Eq.~\eqref{eq:muqzqubit} and find a functional dependence of $q(\mu,u)$ that is given by:
\begin{equation}
q=\frac{\sqrt{u^2+\nu^2+2\nu}- u}{\nu} ,
\end{equation}
where $\nu:= u\mu-(\mu+u)$.
Then, by substituting this expression for $q$ into Eq.~\eqref{eq:fqbitprime1}
we obtain
\begin{equation}
f_{\rm qbit}^\prime (1) =-\frac{\nu [u(u+5)-2]+u\{(2-N_\nu + u[3(u+1)+N_\nu] \}+\nu \{3+u[9-N_\nu +u(3+u+N_\nu)] \} }{2(1-u^2)^2} ,
\end{equation}
\end{widetext}
where $N_\nu = \sqrt{u^2+\nu^2+2\nu}$.
Using a computer algebra system, we can finally maximize the above expression over $0\leq u\leq 1$ for fixed $\mu >2$ and we find the maximum value
\begin{equation}
\max_u \left[ \mu (\mu-2) -2\sigma^2 \right] =\frac{\mu^2 (\mu-2)^2}{2(\mu-1)^2} > 0 ,
\end{equation}
which is obtained for
\begin{equation}
u = \frac{2\mu (\mu-2)}{2\mu^2-4\mu + 4} .
\end{equation}
Substituting back this solution into the above relations between $q$, $u$ and $\mu$ we get 
\begin{equation}
q=1-\frac 2 \mu \qquad \mbox{and} \qquad  u=\frac{2q}{1+q^2} ,
\end{equation}
which is the model described in the main text. With the above substitution, we obtain 
\begin{equation}
{\sigma^2 =\mu(\mu-2)\left[\mu(\mu-2)+2\right]/[4(\mu-1)^2]},
\end{equation}
 and the corresponding expression for the accuracy
\begin{equation}
R=4\mu(\mu-1)^2/\left\lbrace(\mu-2)\left[\mu(\mu-2)+2\right]\right\rbrace,
\end{equation}
giving  $R= 4= d^2$ for $\mu \rightarrow \infty$.
We can also write $U_0$ as ${U_0=\exp(i\theta \sigma_y)}$ where $\sigma_y$ is the Pauli $y$ matrix and $\theta=\arctan\left( \sqrt{1-u}/\sqrt{u}\right)$, which becomes, with the optimal value for $u$ and $q$
\begin{equation}\label{eq:uthetaofq}
\theta = \arctan\left((1-q)/\sqrt{2q} \right)= \arctan\left(\sqrt 2 /\sqrt{\mu(\mu-2)}\right).
\end{equation}
This will be useful later, in Sect.~\ref{app:cont2}, in order to compute the continuous limit.
As a final remark for this section we note that with this relation between $u$ and $q$ we get
\begin{equation}
\Gamma_{q,u} = -\frac{2q(1-q)^2}{1+q^2} ,
\end{equation}
and substituting the optimal value for $q$ we get
\begin{equation}
\begin{aligned}
v_{\pm} &= -(1\pm i)\sqrt{\frac{(\mu-2)}{2\mu}} , \\
\kappa_{\pm} &= (\mu-1\pm i)\sqrt{\frac{(\mu-2)}{\mu(\mu(\mu-2)+2)}} .
\end{aligned}
\end{equation}

\subsection{Generalizations to higher dimensions}\label{sec:qudit_case}

As an outlook to construct a quantum clock model for general dimensions, we can 
now make a series of considerations from what we have learned from the classical case and from the qubit clock model described above.
First of all, by convexity one can fix the initial state to be an arbitrary pure state, say the state $\ket{0}$ of the computational basis, and since we are interested in the sequence $00\ldots01$, it is enough to define only the map $\II_0$, which can be parametrized, e.g., in terms of its Kraus operators $\II_0(\rho)=\sum_{i=1}^{d^2} K_0^{(i)} \rho (K_0^{(i)})^\dagger$. 
However, such a problem is already too complex to be treated, e.g., to perform some basic numerical optimization. Since we are now interested only in obtain a quantum model that violates the classical bound, and not to compute the quantum bound, we can simplify the problem assuming only one Kraus operator $K_0$, i.e., $\II_0(\rho):=K_0 \rho K_0^\dagger$. This is a simplified model and it is not guaranteed to give the maximal quantum value. Even in this simplified scenario, a full optimization is still very difficult to do, even in small dimensions. However, we impose the form of the Kraus operator by analogy with the classical strategy and based on what we learned in the qubit case. Then, one possible solution has the following properties:
\begin{enumerate}
\item There is a single Kraus operator for the outcome ``$0$'', that has the form $K_0=U \sqrt E_0$
\item The initial state is $\ket{\psi}=\ket{0}$ in the computational basis
\item The effect $E_0$ has the diagonal form $E_0={\rm diag}(1,\dots,1,q^2)$ in computational basis
\item The unitary is a one parameter family of the form \begin{equation}
U(\theta)=\exp (i H_{\rm clock}\theta) ,
\end{equation}
where $H_{\rm clock}=F (\sum_k k \ketbra k) F^\dagger$ is obtained by Fourier transforming the computational basis.
\end{enumerate}
Afterwards, there is still to optimize over the parameters $q$ and $\theta$. 
This over-simplified optimization, however, is still too hard to solve for general dimensions. 

\begin{figure}
\includegraphics[width=0.48\textwidth]{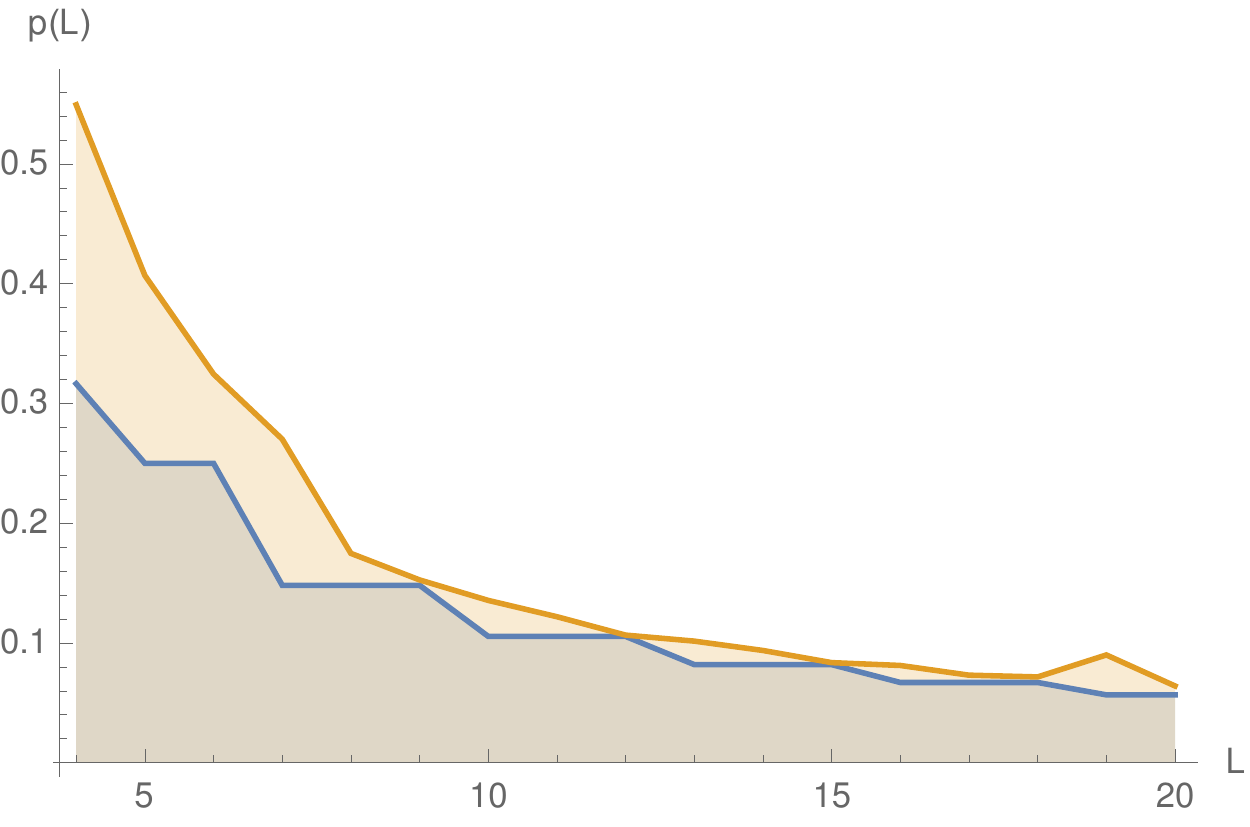}
\includegraphics[width=0.48\textwidth]{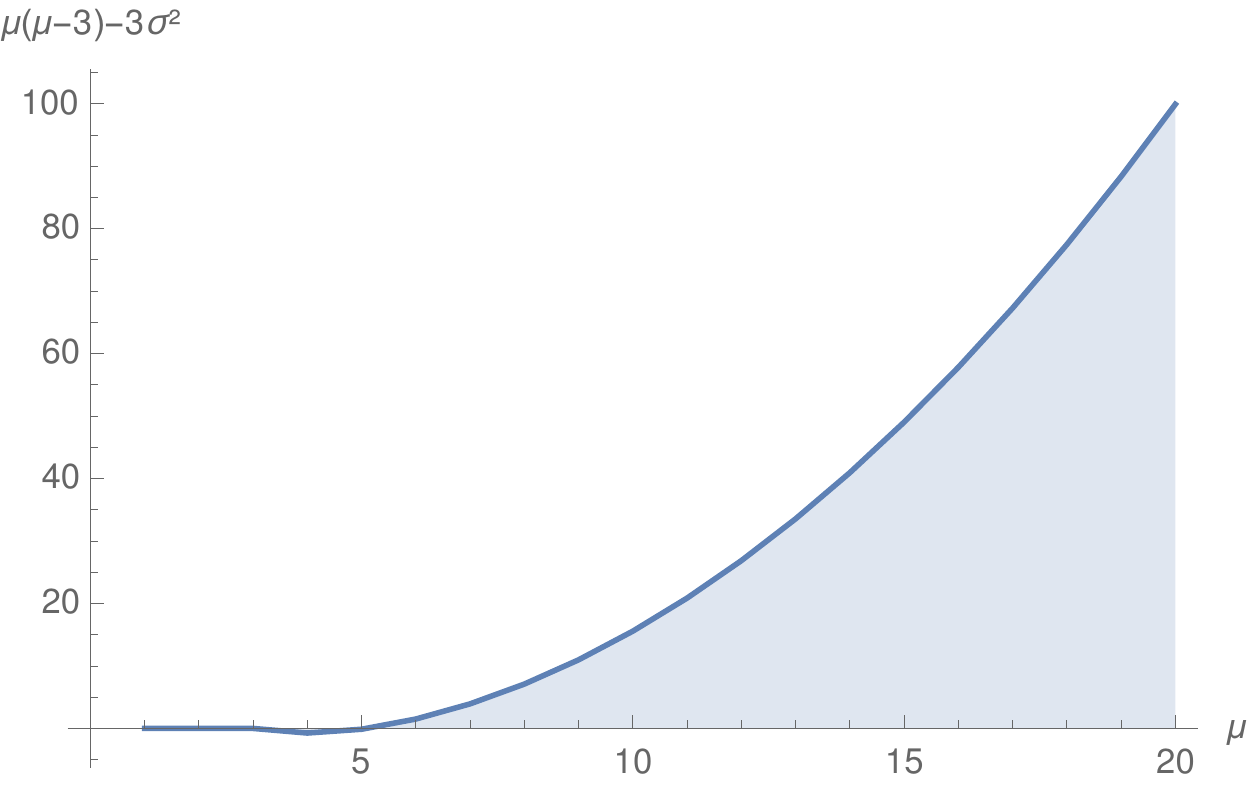}
\caption{\label{fig:plot_Q_Cqutrit} Left panel: plot of $p(L)$ for the best multicyclic model (blue) and the quantum model described here with parameters $u=2 q/(1+q^2)$ and $q=(1-3/L)$ (yellow). Right panel: plot of the left hand side of Eq.~\eqref{eq:conj1} of the main text, i.e., $\mu(\mu-d)-d\sigma^2$ for $d=3$, achieved with the same quantum model with $q=(1-3/\mu)$ and $\mu\geq 4$.}
\end{figure}

\begin{widetext}
In the following, we provide some example of a quantum model beating the (conjectured) classical bound for the case $d=3$. 
To use real parameters we parametrized the unitary as
\begin{equation}
U=\frac 1 3 \left( 
\begin{matrix}
4u-1 & 2(1-u)+2\sqrt{3u(1-u)} & 2(1-u)-2\sqrt{3u(1-u)} \\
2(1-u)-2\sqrt{3u(1-u)} & 4u-1 & 2(1-u)+2\sqrt{3u(1-u)} \\
2(1-u)+2\sqrt{3u(1-u)} & 2(1-u)-2\sqrt{3u(1-u)} & 4u-1
\end{matrix}
\right) ,
\end{equation}
\end{widetext}
where $0\leq u\leq 1$. Numerical results are shown in Fig.~\ref{fig:plot_Q_Cqutrit}, where we plot: (left) the probability $p(L)$ for the quantum model described above, optimized over $q$ and $u$ together with the maximum achieved by the multicyclic models and (right) the value of the expression $3 \sigma^2 - \mu ({\mu-3})$ for the same model, but where
we also fix the functional relation 
between the parameters $q$ and $u$ to be the same as what we found in the qubit case, namely:
\begin{equation}
u = \frac{2q}{1+q^2} ,
\end{equation}
and the value of $q$ as
\begin{equation}
\qquad q=1-d/\mu .
\end{equation}
Thus, we find from numerics that such a model, even if resulting from an over-simplified optimization, already outperforms the classical multicyclic models for $d=3$ in both cases. Thus, we believe that a similar model would work for general dimension.

\section{Continuous limit of quantum clock models}\label{app:cont2}

We now compute explicitly the continuous limit of the quantum model presented in the main text. For an initial pure state and single-Kraus-operator evolution, we write 
$\ket{\psi(t)}=\me^{(iHt-Vt)}\ket{\psi(0)}$, with $H$ Hermitian 
and $V$ positive semidefinite. The Kraus operator is $K_0=U_0\sqrt{E_0}$, so that the discrete evolution is  
$\ket{\psi_n}=K_0^n \ket{\psi_0}$ and we can put $K_0=\me^{\delta(-V+iH))}\approx \openone + \delta (-V+iH)$
for $\delta \approx 0$. From this relation and with
$q=q(\delta)$, we can associate
\begin{equation}
-V+iH = \lim_{\delta\rightarrow 0^+} \frac{K_0 -\openone}{\delta} = \lim_{\delta\rightarrow 0^+} \frac{\log K_0}{\delta} .
\end{equation}
To calculate the logarithm of $K_0$ we can make use of 
the Baker-Campbell-Hausdorff formula and obtain
\begin{equation}
\log K_0 = \tfrac{\log(q(\delta))} 2 (\openone+\sigma_z) + i\theta(q(\delta)) \sigma_y + o(q(\delta)) ,
\end{equation}
and finally, substituting the value of $\theta(q)$ from \eqref{eq:uthetaofq} and 
recalling $q(\delta)\approx 1- \alpha \delta$  to first order, we have
\begin{equation}
\begin{split}
\tfrac 1 \delta \log K_0 \approx \tfrac 1 \delta \left[\tfrac{-\alpha \delta} 2 (\openone+\sigma_z) + i\tfrac{\alpha \delta}{\sqrt 2} \sigma_y \right] = \\
= \alpha \left[(\openone+\sigma_z)/2 +i\sigma_y/\sqrt 2 \right] .
\end{split}
\end{equation}
Thus we can identify $V=(\openone+\sigma_z)/2$ and $H=\sigma_y/\sqrt 2$ and obtain in this way the continuous limit of our quantum model.

\section{Computing lower and upper bounds on maximum classical correlations via grid-search methods}\label{app:low_up_b}

In this section, we show how the maximum classical correlations can be estimated via grid-search methods. Intuitively, since the function we want to maximize is a polynomial defined on a compact set (a cartesian product of simplexes, each associated with the parameters of a row of the transition matrix $T_0$), its gradient will be bounded. Hence, it is possible to obtain an approximation with a bounded error, by evaluating the function only on a finite number of lattice points in the parameter space. 

\subsection{General considerations}
Consider a function $f:\mathcal{B}\subset\mathbb{R}^D\rightarrow \mathbb{R}$, where $\mathcal{B}$ is a compact set, and $f$ is at least $C^1$. Let us consider the following optimization problem

\begin{equation}\label{eq:max_f}
\begin{split}
\text{max: } & f(\xx) \\
\text{subject to: }& \xx \in \mathcal{B}
\end{split}
\end{equation}

Since $f$ and $\|\nabla f\|$ are continuous functions and $\mathcal{B}$ is compact, they have a maximum in $\mathcal{B}$. For $\delta\in \mathbb{R}^+$, let us define $\mathcal{L}_\delta = (\delta\mathbb{Z}^D)\cap \mathcal{B}$, i.e., the intersection of the cubic lattice of step $\delta$ with the compact set $\mathcal{B}$. By construction, we have that each point in $\mathcal{B}$ is at a distance at most $\delta\sqrt{D}/2$ from $ \LL_\delta$, namely,
\begin{equation}
\max_{\xx  \in \mathcal{B}} \min_{\yy \in \LL_\delta} (\| \xx - \yy \| ) = \delta\sqrt{D}/2,
\end{equation}
where $\xx$ corresponds to the intersection of all the diagonals of the $D$-dimensional hypercube with edge of length $\delta$.

We can now prove a simple upper and lower bound for the problem in Eq.~\eqref{eq:max_f}
\begin{equation}\label{eq:lower_upper_f}
\max_{\xx \in \LL_\delta} f(\xx) \leq \max_{\xx \in \mathcal{B}} f(\xx)\leq \max_{\xx \in \LL_\delta} f(\xx) + \frac{\delta\sqrt{D}}{2} \max_{\xx \in \mathcal{B}} \|\nabla f(\xx)\| .
\end{equation}
The proof is straightforward: for the first inequality it is sufficient to use the inclusion $\LL_\delta \subset \mathcal{B}$, whereas for the second it is sufficient to use the $0^{\rm th}$-order Taylor expansion with the Lagrange remainder, i.e.,
\begin{equation}
f(\xx)= f(\yy) + \nabla f(\zz)\cdot (\xx - \yy),
\end{equation}
where $\zz = \lambda \xx + (1-\lambda) \yy$ for some $0\leq \lambda\leq 1$. 

Notwithstanding the generality of this idea, it may work only if, a) the set $\mathcal{B}$ is easily characterized (e.g., a product of simplexes for the classical model), b) the space of parameters is small (e.g., small dimensional machines), and c) the gradient is easily upper bounded (e.g., for a sequence of length $L$ generated by a classical machine the gradient can be interpreted as a vector of probabilities, where each component is bounded by one).

In fact, the number of points of the lattice scales polynomially with the inverse of the error , e.g., if we are inside the $[0,1]^D$ cube and we take $\delta=1/N$, we need to compute $N^D$ points. Moreover, since the error can be evaluated at the beginning (e.g., as $\delta\sqrt{D}L$) and computations on a lattice points are independent of each other, this process can be highly parallelized, e.g., on a GPU.

\subsection{Classical case}
We now specialize the above result to the case of classical $d$-state machines, corresponding to $D=d^2$ parameters. Eq.~\eqref{eq:max_f} for the case of $p_L(T)$ then becomes
\begin{equation}\label{eq:max_T}
\begin{split}
\text{max: } & p_L(T)=\pi T^{L-1}(1-T)\eta \\
\text{subject to: }& T_{ij} \geq 0, \text{ for } i,j,=1,\ldots,d\\
& \sum_{j=1}^d T_{ij} \leq 1,\ \text{ for  } i = 1,\ldots,d.
\end{split}
\end{equation}
We can compute the partial derivate of the above expression as 
\begin{equation}
\frac{\de T^n}{\de T_{ij}} = \sum_{r=0}^{n-1} T^r J^{ij} T^{n-r-1},
\end{equation}
where $J^{ij}$ is the matrix with a $1$ in position $(i,j)$ and $0$ otherwise. It is convenient to represent  using the Dirac notation as $J^{ij}=\ket{i}\bra{j}$,   even though the model is classical. Let us denote also $\pi=\bra{1}$ and $\eta=\ket{\eta}$.
We can then write
\begin{equation}\label{eq:grad_ij}
\begin{split}
&\frac{\de }{\de T_{ij}}\bra{1}T^{L-1}(1-T)\ket{\eta} \\
&= \sum_{r=0}^{L-2}\bra{1} T^r \ket{i}\bra{j}T^{L-r-2}(1-T)\ket{\eta} - \bra{1} T^{L-1} \ket{i}\mean{j | \eta}\\
&= \sum_{r=0}^{L-2}p_{1,i}(0^r) p_{j,-}(0^{L-r-2}1) - p_{1,i}(0^{L-1})
\end{split}
\end{equation}
where $p_{\alpha,\beta}(0^k 1^h)$ is defined as the probability of outputting first $k$ zeros, then $h$ ones, when starting from the state $\alpha$ and ending in the state $\beta$, whereas the notation $p_{j,-}$ denotes a sum over the final state. 

One can prove that the expression in Eq.~\eqref{eq:grad_ij} is smaller than one in absolute value, giving $\max_{T \in \mathcal{B}} \|\nabla p_L(T)\|\leq \sqrt{D}=d$. Moreover, by the geometry of the set $\mathcal{B}$ we can have as a maximal distance $\delta\sqrt{D}/2$, providing 
\begin{equation}\label{eq:lower_upper_T}
\begin{split}
\max_{T \in \LL_\delta} p_L(T) \leq \max_{T \in \mathcal{B}}p_L(T)\leq \max_{T \in \LL_\delta} p_L(T) + \frac{\delta \sqrt{D}}{2}\sqrt{D}\\
= \max_{T \in \LL_\delta} p_L(T) + \delta  d^2/2 .
\end{split}
\end{equation}
Let us prove the above estimate for the gradient. First, notice that, since both terms are positive,
\begin{equation}\label{eq:max2grad}
\begin{split}
&\left| \sum_{r=0}^{L-2}p_{1,i}(0^r) p_{j,-}(0^{L-r-2}1) - p_{1,i}(0^{L-1}) \right| \\
&\leq \max\left\lbrace\sum_{r=0}^{L-2}p_{1,i}(0^r) p_{j,-}(0^{L-r-2}1) ,\  p_{1,i}(0^{L-1})\right\rbrace.
\end{split}
\end{equation}
We can then proceed to bound each term separately. Clearly $p_{1,i}(0^{L-1})\leq 1$, since it is a probability. Actually, one can also prove that $\sum_i p_{1,i}(0^{L-1})=p_{1,-}(0^{L-1})\leq 1$. For the other term, we just notice that
\begin{equation}
\begin{split}
\sum_{r=0}^{L-2}p_{1,i}(0^r) p_{j,-}(0^{L-r-2}1)\leq \sum_{r=0}^{L-2} p_{j,-}(0^{L-r-2}1)\\
\leq \sum_{r=0}^\infty p_{j,-}(0^r1)\leq 1,
\end{split}
\end{equation}
where we used that $p_{1,i}(0^r)\leq 1$ and we identified $\sum_{r=0}^\infty p_{j,-}(0^r1)$ with the probability of occurrence of the output $1$ for a given machine starting in the state $j$, which is either $0$, when the outcome $1$ never appears, or $1$. Again, notice that we obtain that also the sum over $i$, i.e., $\sum_i \sum_{r=0}^{L-2}p_{1,i}(0^r) p_{j,-}(0^{L-r-2}1)$, is smaller than $1$. Since each term $\de_{ij}p_L(T)\leq 1$, we obtain $\|\nabla p_L(T)\|\leq \sqrt{D}=d$.

\begin{table}[t]
\begin{tabularx}{0.5\textwidth}{ | X | X | X | X |}
\hline
\multicolumn{4}{|c|}{Q max vs. C estimate and upper bound, $d=2$ }\\
\hline
$\qquad L \qquad $ & $p_{\max}^{\rm Num\ Q}$  & $p_{\max}^{\rm Est\ C}$ & $p_{\max}^{\rm UB\ C}$  \\
 \hline
3 & 0.3792   & 0.2963 & 0.29641\\
4 & 0.2525 & 0.25 & 0.2501  \\
5 & 0.1907 & 0.14815 & 0.1483\\
6 & 0.1528 & 0.14815 & 0.1483\\
7 & 0.1274 &  0.105469 & 0.1056 \\ 
8 & 0.1093 &  0.105469 & 0.1056 \\ 
9 & 0.0957 &  0.08192 & 0.0821\\
10& 0.0851 &  0.08192 & 0.08195\\
11 & 0.07659 & 0.0669796 & 0.06701\\
12 & 0.06963 & 0.0669796 & 0.06701\\
13 & 0.06384 &0.0566528 & 0.05668\\
14 & 0.05893 & 0.0566528 & 0.05668\\
15 & 0.05473 & 0.049087 & 0.04912\\
16 & 0.05107 & 0.049087 & 0.04912 \\
17 & 0.04790 & 0.0433049 & 0.04333 \\
18 & 0.04508 & 0.0433049 & 0.04333 \\
19 & 0.04258 & 0.038742 & 0.03877\\
20 & 0.04034 & 0.038742 & 0.03877\\
\hline
\end{tabularx} 
\caption{\label{tab:QPCA} Numerical values obtained for classical and quantum models of dimension 2:  $p_{\max}^{\rm Num\ Q}$ is the value obtained numerically for the quantum case with the explicit model described in Sec.~\ref{sec:qubit_case}; $p_{\max}^{\rm Est\ C}$ is the estimated optimal solution coming from the cyclic model, except for $L=3$ where the one-way model is better; $p_{\max}^{\rm UB\ C}$ is the analytical upper bound for the classical case computed via lattice calculations.}

\end{table}

Given the above estimate of the gradient, we can compute an upper bound on the maximum of the problem in Eq.~\eqref{eq:max_T} by calculating the value of $p_L(T)$ on a lattice in the space of parameters with lattice step $\delta$. Moreover, since in our case we already have a guess of the optimal solution, such a computation can be reduced by an iterative method that evaluate the function on more and more refined lattices, defined as follows:
\begin{itemize}
\item[$(i)$] We start with an estimated optimal value $p_{\max}^{\rm Est\ C}$, a lattice step $\delta$ and the corresponding lattice $\LL_{\delta}$, and an error $\epsilon(\delta):= \delta d^2$ (we omit the factor $1/2$ for reason that will be clear later). For each $T \in \LL_{\delta}$ if $p_L(T)+\epsilon(\delta) < p_{\max}$, we remove the point from the lattice. After this procedure, we obtain a new lattice $\LL'_{\delta}$. If for some $T$, $p_L(T) >  p_{\max}^{\rm Est\ C}$, we update $p_{\max}^{\rm Est\ C}$ with the this new value.
\item[$(ii)$] For each of the remaining points in  $\LL'_{\delta}$, create a new lattice of length $\delta$ in each direction and lattice step $\delta^2$. Let us denote this lattice as $\LL^1_{\delta^2}$ and the corresponding error $\epsilon(\delta^2)$. Again, for each $T\in \LL^1_{\delta^2}$, if $p_L(T)+\epsilon(\delta^2) < p_{\max}^{\rm Est\ C}$, we remove the point from the lattice. We obtain the new lattice $\LL_{\delta^2}^{'1}$ and we update $p_{\max}^{\rm Est\ C}$ if a better estimate is found.
\item[$(iii)$] Iterate the procedure until the desired error $\epsilon$
is reached.
\end{itemize}

Notice that, with the exception of the last iteration, the factor $1/2$ for the error estimate cannot be used as a consequence of the way we construct the refined lattice. Such a procedure is illustrated for the simple one-dimensional case in Fig.~\ref{fig:iter}.

With the above iterative method, we were able to evaluate $\max p_L(T)$ for $d=2$ and $L=3,4,\ldots,9$ over a lattice of $10^{18}$ points, and $L=10,11,\ldots,20$ over a lattice of $10^{19}$ points. As a result, we can certify that the one-way model for $L=3$ and the cyclic model for $L=4,\ldots,20$ provide the optimal value for the problem in Eq.~\eqref{eq:max_T}, up to an error of $10^{-4}$ for $L=3, \ldots, 9$  and of $3\times 10^{-5}$ for $L=10,\ldots, 20$. The results are collected in Tab.~\ref{tab:QPCA}.

\begin{table*}[t!]

\begin{tabularx}{\textwidth}{| X | X | X | X | X | X | X | X | X | X | X |}
\hline
\multicolumn{11}{|c|}{Optimum found by numerical search for given $d$ and $L$}\\
\hline
\diagbox[width=3em]{$\ d$}{ \vphantom{$\frac{L}{L}$} $L \ $} & d+1 & d+2 & d+3 & d+4 & d+5 & d+6 & d+7 & d+8 & d+9 & d+10\\
\hline
3 &  F &  F &  F &  F &  28\% &  39\% &  22\% &  32\% &  39\% &  39\%\\
\hline
4 &  F &  F &  F &  25\% &  22\% &  32\% &  40\% &  46\% &  32\% &  32\%\\
\hline
5 &  F &  19\% &  30\% &  42\% &  51\% &  28\% &  36\% &  42\% &  47\% &  47\%\\
\hline
6 &  F &  26\% &  39\% &  40\% &  48\% &  55\% &  32\% &  39\% &  44\% &  44\%\\
\hline
7 &  F &  5\% &  25\% &  38\% &  47\% &  53\% &  57\% &  36\% &  42\% &  42\%\\
\hline
8 &  F &  26\% &  36\% &  46\% &  45\% &  52\% &  56\% &  61\% &  39\% &  39\%\\
\hline
9 &  F &  23\% &  39\% &  35\% &  44\% &  50\% &  55\% &  60\% &  63\% &  63\%\\
\hline
10  &  F &  26\% &  34\% &  44\% &  52\% &  49\% &  54\% &  59\% &  62\% &  62\%\\
\hline
\end{tabularx}
\caption{\label{tab:Adam} Results of the numerical search for $d=3,\ldots,10$ and $L=d+1,\ldots,d+10$. The letter F denotes that the optimal model has been found by the numerical search, up to the a numerical precision of $10^{-5}$. In all other cases, the numerical search found a worse result, the ratio between the gap and the optimal value is indicated in percentage, e.g., for $d=3$ and $L=8$ the algorithm found an optimum which has a gap of $28\%$ with respect to the optimum of the cyclic ($k=3$) model, i.e., $({\rm opt}_{\rm cy} - {\rm opt}_{\rm Adam})/ {\rm opt}_{\rm cy}  = 28\%$.}
\end{table*}

\section{Numerical search for inequalities in higher dimension}\label{app:num_search}

In this section, we discuss the numerical search for violation of the generalized inequalities 
\begin{eqnarray}\label{eq:conj_App_1}
 d \sigma^2 - \mu ({\mu-d}) \geq 0,\\
p(L)\leq \Omega_{d,L}^C . \label{eq:conj_App_2}
\end{eqnarray}
To perform the optimization, we used the Adam algorithm~\cite{Adam} implementation present in the {\it pytorch} package~\cite{Pytorch}. 
The space of parameters for our optimization consists in the space of substochastic matrices, which are constrained by positivity and normalization conditions. We first transform the problem into an unconstrained one. For a given dimension $d$, we construct our transition matrices as follows. Let $B_0, B_1$ be two $d \times d$ matrices with real coefficients. We define
\begin{equation}
[T_k]_{ij} := \frac{([B_k]_{ij})^2}{\sum_j \left[([B_0]_{ij})^2 + ([B_1]_{ij})^2\right]}, \text{ for } k=0,1 ,
\end{equation}
such that $T_0$ and $T_1$ are substochastic matrices and $T_0+T_1$ is stochastic, for any choices of $B_0, B_1$. This transforms the constrained optimization problem into an unconstrained one, which can be attacked with the Adam algorithm. We fix the parameter of the optimization, i.e., number of steps, ``learning rate'', i.e., the size of each step, and number of repetitions of the optimization after few simple tests. We tried to perform the same optimization with the stochastic gradient descent algorithm, but obtained worse results.

\subsection{Accuracy}

By using Eq.~\eqref{eq:muqzcl}, we have
\begin{equation}\label{eq:conj1_expanded}
\begin{split}
d \sigma^2 - \mu ({\mu-d}) &= 2 d \pi_1  ( \openone -  T_0 )^{-2}\eta \\
 &- (d+1) (\pi_1  ( \openone -  T_0 )^{-1}\eta)^2\geq 0.
\end{split}
\end{equation}
We recall that $\left( \openone -  T_0 \right)^{-1}=\frac{1}{{\rm det}(\openone- T_0)}{\rm adj}(\openone- T_0)$, where ${\rm adj}$ denotes  the adjugate matrix (cf. Appendix.~\ref{app:2_dim_cl}). Since we are interested only in the positivity of the expression in Eq.~\eqref{eq:conj1_expanded}, we can multiply it by $({\rm det}(\openone- T_0))^2$. Hence, we obtain the expression  
\begin{equation}\label{eq:def_FB}
F[B_0,B_1] := 2 d \pi_1  \left( {\rm adj}[\openone -  T_0] \right)^{2}\eta - (d+1) (\pi_1  {\rm adj}\left[ \openone -  T_0 \right]\eta)^2 .
\end{equation}
To find a violation of Eq.~\eqref{eq:conj_App_1}, we perform the minimization
\begin{equation}
\begin{split}
\text{Minimize: }& F[B_0,B_1] \\
\text{subject to: }& [B_0]_{ij}, [B_1]_{ij}\in \mathbb{R}, \text{ for all } i,j=1,\ldots,d. 
\end{split}
\end{equation}
The package pytorch automatically computes the gradient of the expression in Eq.~\eqref{eq:def_FB} and performs the optimization. For each $d\in \{3,\ldots,10\}$, we performed 100 times the optimization starting from a random initial point and with $10^4$ optimization steps and a ``learning rate'' of $0.005$. For all dimensions, the optimization converges to a value of $10^{-5}$, i.e., approximately $0$, consistent with the conjectured inequality and optimality of the one-way model.  

\subsection{Finite sequences}
Using again the parametrization of $T_0,T_1$ in terms of $B_0,B_1$ we have 
\begin{equation}
G[B_0,B_1] = \pi_1 T_0^{L-1}(\openone- T_0)\eta
\end{equation}
and the problem
\begin{equation}
\begin{split}
\text{Maximize: }& G[B_0,B_1] \\
\text{subject to: }& [B_0]_{ij}, [B_1]_{ij}\in \mathbb{R}, \text{ for all } i,j=1,\ldots,d. 
\end{split}
\end{equation}
which we optimize again via Adam for $d\in \{ 3,\ldots, 10\}$ and $L \in \{d+1,\ldots, d+10\}$. For each pair $(d,L)$ we repeat the optimization $100$ times with a randomly generated initial point, $10^4$ steps for each optimization, and a learning rate of $0.005$.
Typically, Adam is able to find the correct value for low dimension and short sequences, as it was to be expected. In no case the algorithm found a better value than those already known. The results are summarized in Table~\ref{tab:Adam}.